\documentclass[apj]{emulateapj}
\usepackage{apjfonts}
\usepackage{amsbsy}

\def\hmpc{$h^{-1}$Mpc}
\def\hkpc{$h^{-1}$kpc}

\def\mstar{$M_\ast$}

\def\hmsol{$h^{-1}$M$_\odot$}

\def\om{\Omega_m}

\def\s8{\sigma_8}

\def\lcdm{$\Lambda$CDM}

\def\x2{$\chi^2$}
\def\hmsol{$h^{-1}\,$M$_\odot$}

\def\NNm1{\langle N(N-1) \rangle}

\def\m_star{M_\ast}

\def\lcdm{$\Lambda$CDM}

\def\om{\Omega_m}

\def\s8{\sigma_8}

\def\hmpc{$h^{-1}\,$Mpc}
\def\hkpc{$h^{-1}\,$kpc}

\def\x2{$\chi^2$}
\def\hmsol{$h^{-1}\,$M$_\odot$}

\def\mstar{M_\ast}

\def\NNm1{\langle N(N-1) \rangle}

\def\mstar{M_\ast}

\def\p0{P_0(r)}

\def\siginv{\sigma^{-1}}
\def\fsig{f(\sigma)}
\def\gsig{g(\sigma)}
\def\D{\Delta}
\def\rd{R_{\Delta}}
\def\xdof{\chi^2/\nu}

\def\rhobar{\bar{\rho}_m}
\def\link{{l}}

\begin{document}

\title{Toward a halo mass function for precision cosmology:\\
  the limits of universality}

\author{
Jeremy Tinker\altaffilmark{1,2}, 
Andrey V. Kravtsov\altaffilmark{1,2,3},
Anatoly Klypin\altaffilmark{4},
Kevork Abazajian\altaffilmark{5},\\
Michael Warren\altaffilmark{6},
Gustavo Yepes\altaffilmark{7},
Stefan Gottl{\"o}ber\altaffilmark{8},
Daniel E. Holz\altaffilmark{6}
}
\altaffiltext{1}{Kavli Institute for Cosmological Physics, The University of Chicago, 5640 S. Ellis Ave., Chicago, IL 60637, USA}
\altaffiltext{2}{Department of Astronomy \& Astrophysics, The University of Chicago, 5640 S. Ellis Ave., Chicago, IL 60637, USA}
\altaffiltext{3}{Enrico Fermi Institute, The University of Chicago, 5640 S. Ellis Ave., Chicago, IL 60637, USA}
\altaffiltext{4}{Department of Astronomy, New Mexico State University}
\altaffiltext{5}{Department of Physics, University of Maryland, College Park}
\altaffiltext{6}{Theoretical Astrophysics, Los Alamos National Labs} 
\altaffiltext{7}{Grupo de Astrofísica, Universidad Autónoma de Madrid}
\altaffiltext{8}{Astrophysikalisches Institut Potsdam, Potsdam, Germany}


\begin{abstract}

  We measure the mass function of dark matter halos in a large set of
  collisionless cosmological simulations of flat $\Lambda$CDM
  cosmology and investigate its evolution at $z\lesssim 2$. Halos are
  identified as isolated density peaks, and their masses are measured
  within a series of radii enclosing specific overdensities. We argue
  that these spherical overdensity masses are more directly linked to
  cluster observables than masses measured using the
  friends-of-friends algorithm (FOF), and are therefore preferable for
  accurate forecasts of halo abundances. Our simulation set allows us
  to calibrate the mass function at $z=0$ for virial masses in the
  range $10^{11}$ \hmsol\ $\le M\le 10^{15}$ \hmsol\, to $\lesssim
  5\%$. We derive fitting functions for the halo mass function in this
  mass range for a wide range of overdensities, both at $z=0$ and
  earlier epochs. In addition to these formulae, which improve on
  previous approximations by $10$-$20\%$, our main finding is that the
  mass function cannot be represented by a universal fitting function
  at this level of accuracy. The amplitude of the ``universal''
  function decreases monotonically by $\approx 20-50\%$, depending on
  the mass definition, from $z=0$ to $2.5$. We also find evidence for
  redshift evolution in the overall shape of the mass function.
\end{abstract}



\section{Introduction}

Galaxy clusters are observable out to high redshift ($z\lesssim
1$--$2$), making them a powerful probe of cosmology. The large numbers
and high concentration of early type galaxies make clusters bright in
optical surveys, and the high intracluster gas temperatures and
densities make them detectable in X-ray and through the
Sunyaev-Zel'dovich (SZ) effect. The evolution of their abundance and
clustering as a function of mass is sensitive to the power spectrum
normalization, matter content, and the equation of state of the dark
energy and, potentially, its evolution
\citep[e.g.,][]{holder_etal01,haiman_etal01,weller_etal02,majumdar_mohr03}.
In addition, clusters probe the growth of structure in the Universe,
which provides constraints different from and complementary to the
geometric constraints by the supernovae type Ia
\citep[e.g.,][]{albrecht_etal06}.  In particular, the constraints from
structure growth may be crucial in distinguishing between the
possibilities of the cosmic acceleration driven by dark energy or
modification of the magnitude-redshift relation due to the non-GR
gravity on the largest scales \citep[e.g.,][]{knox_etal05}.

The potential and importance of these constraints have motivated
current efforts to construct several large surveys of high-redshift
clusters both using the ground-based optical and Sunyaev-Zel'dovich
(SZ) surveys and X-ray missions in space.  In order to realize the
full statistical power of these surveys, we must be able to make
accurate predictions for abundance evolution as a function of
cosmological parameters.

Traditionally, analytic models for halo abundance as a function of
mass, have been used for estimating expected evolution
\citep{press_schechter:74,bond_etal91,lee_shandarin:98,sheth_tormen:99}. Such
models, while convenient to use, require calibration against
cosmological simulations.  In addition, they do not capture the entire
complexity of halo formation and their ultimate accuracy is likely
insufficient for precision cosmological constraints. A precision mass function
can most directly be achieved through explicit cosmological simulation.

The standard for precision determination of the mass function from
simulations was set by \cite{jenkins_etal:01} and
\citet{evrard_etal:02}, who have presented fitting function for the
halo abundance accurate to $\sim 10-20\%$. These studies also showed
that this function was universal, in the sense that the same function
and parameters could be used to predict halo abundance for different
redshifts and cosmologies. \cite{warren_etal:06} have further improved
the calibration to $\approx 5\%$ accuracy for a fixed cosmology at $z=0$. 
Several other studies have tested the universality of the
mass function at high redshifts (\citealt{reed_etal:03,
reed_etal:07, lukic_etal:07, cohn_white:07}).

One caveat to all these studies is that the theoretical counts as a
function of mass have to be converted to the counts as a function of
the cluster properties observable in a given survey. Our understanding
of physics that shapes these properties is, however, not sufficiently
complete to make reliable, robust predictions. The widely adopted
strategy is therefore to calibrate abundance as a function
of total halo mass and calibrate the relation between mass and
observable cluster properties either separately or within a survey
itself using nuisance parameters
\citep[e.g.][]{majumdar_mohr04,lima_hu04,lima_hu05,lima_hu07}. 
The success of such a strategy, however, depends on how well cluster observables
correlate with total cluster mass and whether evolution of this
correlation with time is sufficiently simple
\citep[e.g.,][]{lima_hu05}.

Tight intrinsic correlations between X-ray, SZ, and optical
observables and cluster mass are expected theoretically
\citep[e.g.,][]{bialek_etal:01,dasilva_etal04,motl_etal05,nagai06,kravtsov_etal06}
and were shown to exist observationally
\citep[e.g.,][]{mohr_etal:99,lin_etal:04a,vikhlinin_etal06,maughan07,arnaud_etal07,sheldon_etal:07_data,zhang_etal08}
in the case when both observables and masses are defined within a
certain {\it spherical} radius enclosing a given overdensity. The
majority of the mass function calibration studies, however, have
calibrated the mass function with halos and masses measured using the
friends-of-friends (FOF) percolation algorithm.  This algorithm is
computationally efficient, straightforward to implement, and is
therefore appealing computationally. The relation between the FOF
masses and observables is, however, quite uncertain.

As we show below (see \S~\ref{sec:fof_vs_so} and Fig.~\ref{fof_SO}),
there is large, redshift-dependent, and asymmetric scatter between the
FOF mass and mass measured within a spherical overdensity, which
implies that there is also large asymmetric scatter between the FOF
mass and cluster observables. This does not bode well for
self-calibration of such relations. Furthermore, there is no way to
measure the equivalent of the FOF mass in observations, which means
that any calibration of the FOF mass and observables will have to rely
on simulations. An additional issue is that halos identified with
an FOF algorithm may not have one-to-one correspondence to the
objects identified in observational surveys. For example, the FOF
finder is known to join neighboring halos into a single object even 
if their centers are located outside each others virial radii. Such
objects, however, would be identified as separate systems in X-ray and SZ 
surveys. 

Although no halo-finding algorithm applied on simulations containing
only dark matter may be perfect in identifying all systems that would
be identified in a given observational survey, the spherical
overdensity (SO) halo finder, which identifies objects as spherical
regions enclosing a certain overdensity around density peaks
\citep{lacey_cole:94}, has significant benefits over the FOF both
theoretically and observationally. Most analytic halo models
\citep[see, e.g.,][for review]{cooray_sheth:02} assume that halos are
spherical, and the statistics derived are sensitive to the exact halo
definition. To be fully self-consistent, the formulae for halo
properties, halo abundance, and halo bias, on which the calculations
rely, should follow the same definition.  The tight correlations
between observables and masses defined within spherical apertures
means that connecting observed counts to theoretical halo abundances
is relatively straightforward and robust. At the same time, the
problem of matching halos to observed systems is considerably less
acute for halos identified around density peaks, compared to halos
identified with the FOF algorithm.

Thus there is substantial need for a recalibration of the halo mass
function based on the SO algorithm for a range of overdensities probed
by observations and frequently used in theoretical calculations ($\sim 200-2000$). 
Such calibration for the standard $\Lambda$CDM cosmology is the main focus
of this paper. Specifically, we focus on accurate calibration of halo abundances for
intermediate and high-mass halos ($\sim 10^{11}-10^{15}h^{-1}\,\rm M_{\odot}$) over the
range of redshifts ($z\sim 0-2$) most relevant for the current and upcoming
large cluster surveys. 

The paper is organized as follows. In \S\ 2 we describe our
simulation set and SO algorithm. In \S\ 3 we present results for the
mass function, demonstrating how our results depend on cosmology and
redshift.  In \S\ 4 we summarize and discuss our results.

Throughout this paper {\it we use masses defined within radii enclosing
a given overdensity with respect to the mean density of the Universe
at the epoch of analysis.}

\begin{figure*}
\epsscale{1.0} 
\plotone{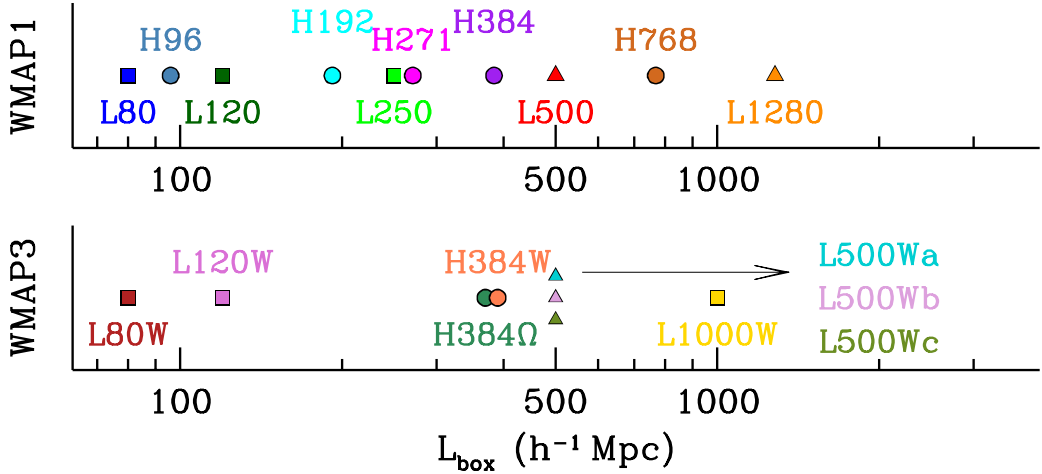}
\vspace{-7.0cm}
\caption{ \label{key} A graphical key for the list of simulations in
  Table 1. The upper panel shows point styles for all the WMAP1 simulations ordered by the
  box size. Each simulation is represented with a different color,
  while different point types represent different numerical codes: circles=HOT,
  squares=ART, triangles=GADGET2. The lower panel plots all
  WMAP3 simulations, as well as H384$\Omega$, the low-$\om$
  simulation. See Table 1 for the details of each simulation. }
\end{figure*}



\begin{deluxetable*}{cccclccclc}
\tablecolumns{7} 
\tablewidth{40pc} 
\tablecaption{Properties of the Simulation Set}
\tablehead{\colhead{$L_{\rm box}$ \hmpc} & \colhead{Name} &  \colhead{$\epsilon$ \hkpc} & \colhead{$N_p$} &\colhead{$m_p$ \hmsol} & \colhead{$(\om,\Omega_b,\s8,h,n)$} & \colhead{Code} & \colhead{$z_i$} & \colhead{$z_{\rm out}$} & \colhead{$\D_{\rm max}$} }

\startdata

768 & H768 & 25 & $1024^3$ &$3.51\times 10^{10}$ & $(0.3,0.04,0.9,0.7,1)$ & HOT& 40 & 0 & 800\\
384 & H384 & 14 & $1024^3$ &$4.39\times 10^{9}$ & $(0.3,0.04,0.9,0.7,1)$ & HOT& 48 & 0 & 3200\\
271 & H271 & 10 & $1024^3$ &$1.54\times 10^{9}$ & $(0.3,0.04,0.9,0.7,1)$ & HOT& 51 & 0 &  3200\\
192 & H192 & 4.9 & $1024^3$ &$5.89\times 10^{8}$ & $(0.3,0.04,0.9,0.7,1)$ & HOT& 54 & 0 &  3200\\
96 & H96 & 1.4 & $1024^3$ &$6.86\times 10^{7}$ & $(0.3,0.04,0.9,0.7,1)$ & HOT& 65 & 0 & 3200\\
\hline
1280 & L1280 & 120 & $640^3$ &$5.99\times 10^{11}$ & $(0.27,0.04,0.9,0.7,1)$ & GADGET2 & 49 & 0, 0.5, 1.0 & 600\\
500 & L500 & 15 & $1024^3\times$2 & $8.24\times 10^{9}$ & $(0.3,0.045,0.9,0.7,1)$ & GADGET2 & 40& 0, 0.5, 1.25, 2.5 & 3200\\
250 & L250 & 7.6 & $512^3$ & $9.69\times 10^{9}$ & $(0.3,0.04,0.9,0.7,1)$ & ART & 49 & 0, 0.5, 1.25, 2.5 & 3200 \\
120 & L120 & 1.8 & $512^3$ & $1.07\times 10^{9}$ & $(0.3,0.04,0.9,0.7,1)$ & ART & 49 & 0, 0.5, 1.25, 2.5 & 3200 \\
80 & L80 & 1.2 & $512^3$ & $3.18\times 10^{8}$ & $(0.3,0.04,0.9,0.7,1)$ & ART & 49 & 0, 0.5, 1.25, 2.5 & 3200\\
\hline
1000 & L1000W & 30 & $1024^3$ & $6.98\times 10^{10}$ & $(0.27,0.044,0.79,0.7,0.95)$ & ART& 60 & 0, 0.5, 1.25, 2.5 & 3200\\
500 & L500Wa & 15 & $512^3\times$2 & $6.20\times 10^{10}$ & $(0.24,0.042,0.75,0.73,0.95)$ & GADGET2& 40 & 0 &  3200\\
500 & L500Wb & 15 & $512^3\times$2 & $6.20\times 10^{10}$ & $(0.24,0.042,0.75,0.73,0.95)$ & GADGET2& 40 & 0  & 3200\\
500 & L500Wc & 15 & $512^3\times$2 & $6.20\times 10^{10}$ & $(0.24,0.042,0.8,0.73,0.95)$ & GADGET2& 40 & 0 & 3200\\
384 & H384W & 14 & $1024^3$ & $3.80\times 10^{9}$ & $(0.26,0.044,0.75,0.71,0.94)$ & HOT& 35 & 0 & 3200 \\
384 & H384$\om$ & 14 & $1024^3$ & $2.92\times 10^{9}$ & $(0.2,0.04,0.9,0.7,1)$& HOT& 42 & 0 &  3200\\
120 & L120W & 0.9 & $1024^3$ & $1.21\times 10^8$ & $(0.27,0.044,0.79,0.7,0.95)$ & ART& 100 & 1.25, 2.5 & 3200\\
80 & L80W & 1.2 & $512^3$ & $2.44\times 10^{8}$ & $(0.23,0.04,0.75,0.73,0.95)$ & ART& 49 & 0, 0.5, 1.25, 2.5 & 3200\\

\enddata \tablecomments{ The top set of 5 simulations are from the
  \cite{warren_etal:06} study. The second list of 5 simulations are of
  the same WMAP1 cosmology, but with different numerical codes. The
  third list of 8 simulations are of alternate cosmologies, focusing
  on the WMAP3 parameter set. The HOT code employs Plummer softening,
  while GADGET employs spline softening. The force resolution of the
  ART code is based on the size of the grid cell at the highest level
  of refinement. $\D_{\rm max}$ is the highest overdensity for which
  the mass function can measured directly. Above this $\D$, halo mass
  are inferred from the rescaling procedure in \S 2.3. A graphical key
  of this table is shown in Figure \ref{key}.}
\end{deluxetable*}

\section{Methods}

\subsection{Simulation Set}

Table 1 lists our set of simulations. All the simulations are based on
variants of the flat, \lcdm\ cosmology. The cosmological parameters
for the majority of the simulations reflect the zeitgeist of the
first-year WMAP results (\citealt{spergel_etal:03}). We will refer to
this cosmology as WMAP1. A smaller number of simulations have 
cosmological parameters closer to the three-year WMAP constraints
(\citealt{spergel_etal:07}), in which both $\om$ and $\s8$ are lower
and the initial power spectrum contains significant tilt of
$n=0.95$. This subset of simulations are not of the same identical
parameter set, but rather represent slight variations around a
parameter set we will refer to globally as WMAP3.

The largest simulation by volume followed a cubic box of 1280 \hmpc\
size. There are fifty realizations of this simulation performed with
the GADGET2 code \citep{springel:05}, which have been kindly provided
to us by R. Scoccimarro. With the exception of these 1280 \hmpc\
boxes, the initial conditions for all simulations were created using
the standard first-order Zel'dovich approximation
(ZA). \cite{crocce_etal:06} point out possible systematic errors in
the resulting mass function if first-order initial conditions are
started insufficiently early. Using second order Lagrange perturbation
theory (2LPT) to create initial conditions, they identify
discrepancies between the halo mass function from their simulations
and that of \cite{warren_etal:06} at the highest masses. In
\cite{warren_etal:06}, several boxes larger than 768 \hmpc\ were
utilized in the analysis that are not listed in Table 1. In tests with
our spherical overdensity halo finder, we find a discrepancy between
the 2LPT simulations and these simulations.  At this point, it is not
yet clear whether the discrepancy is due to the effect advocated by
\cite{crocce_etal:06} or due to other numerical effects. We explore
the issue of initial starting redshift in some detail in the Appendix
A. What is clear, however, is that results of these simulations
systematically deviate from other higher resolution simulations,
especially for larger values of overdensities.  We therefore do not
include them in our analyses.

The first five simulations listed in Table 1 were used 
in \cite{warren_etal:06} in their analyses.  The integrations were performed
with the Hashed Oct-Tree (HOT) code of
\cite{warren_salmon:93}. Additionally, there are two HOT simulations
in the WMAP3 parameter set. These simulations will be referred to in
the text by their box size, in \hmpc, prefixed by the letter
`H'. Simulations in the WMAP3 set will be appended with the letter
'W'. Due to identical box sizes between parameter sets, H384 will
refer to the WMAP1 simulation, H384W will refer to the simulation with
WMAP3 parameters, and H384$\Omega$ will refer to the low-$\Omega_m$
simulation (which we will include in the WMAP3 simulation subset).

There are six simulations using the Adaptive Refinement Technique
(ART) of \cite{kravtsov_etal:97}, and four that use GADGET2 in
addition to the L1280 realizations. The L80 and L120 ART boxes
modeling the WMAP1 cosmology are described in \cite{kravtsov_etal:04}
and L250 simulation is described by Tasitsiomi et al.~(2008, in
preparation), while the three WMAP3 boxes are presented here. The L500
simulations are described in \cite{gottloeber_yepes:07} and
\cite{yepes_etal:07}\footnote{see also
  http://astro.ft.uam.es/marenostrum/universe/index.html}.  These
simulations contain equal numbers of dark matter and SPH gas particles
(without cooling). The ART and GADGET2 simulations will be referred by
their box size with prefix `L'. WMAP3 simulations have a `W' as a
suffix.

Our simulation set comprises three different N-body codes, one based
on the popular tree algorithm (HOT), one based on grid codes with
small-scale refinement of high-density regions (ART), and one that
combines grid and tree algorithms (GADGET2). We present a key in
Figure \ref{key} that graphically displays the range of box
sizes. Each simulation is represented by a different color, while
different point types refer to different simulation codes: circles for
HOT, squares for ART, and triangles for GADGET2. These point symbols
and colors will be used consistently in the figures below.

\subsection{Halo Identification}

The standard spherical overdensity algorithm is described in detail in
\cite{lacey_cole:94}. However, in our approach we have made several
important modifications.  In \cite{lacey_cole:94} the centers of halos
are located on the center of mass of the particles within the
sphere. Due to substructure, this center may be displaced from the
main peak in the density field. Observational techniques such as X-ray
cluster identification locate the center of the halo at the peak of
the X-ray flux (and therefore the peak of density of the hot
intracluster gas). Optical cluster searches will often locate the
cluster center at the location of the brightest member, which is also
expected to be located near the peak of X-ray emission (\citealt{lin_etal:04a,
koester_etal:07, rykoff_etal:08}). Thus we locate the centers of halos at their
density peaks.

Our halo finder begins by estimating the local density around each
particle within a fixed top-hat aperture with radius approximately
three times the force softening of each simulation. Beginning with the
highest-density particle, a sphere is grown around the particle until
the mean interior density is equal to the input value of $\D$, where
$\D$ is the overdensity within a sphere of radius $\rd$ with respect
to the mean density of the Universe at the epoch of analysis,
$\rhobar(z)\equiv\om(z)\rho_{\rm crit}(z)=\rhobar(0)(1+z)^3$:
\begin{equation}
\D = \frac{M_{\Delta}}{(4/3)\pi \rd^3 \rhobar}.
\end{equation}
All values of $\D$ listed in this paper are with respect to $\rhobar(z)$. 

Since local densities smoothed with a top-hat kernel are somewhat noisy,
we refine the location of the peak of the halo density with an
iterative procedure. Starting with a radius of $r=\rd/3$, the center
of mass of the halo is calculated iteratively until convergence. The
value of $r$ is reduced iteratively by 1\% and the new center of mass
found, until a final smoothing radius of $\rd/15$, or until only 20
particles are found within the smoothing radius. At this small
aperture, the center of mass corresponds well to the highest density
peak of the halo. This process is computationally efficient and
eliminates noise and accounts for the possibility that the chosen
initial halo location resides at the center of a large substructure;
in the latter case, the halo center will wander toward the larger mass
and eventually settle on its center. Once the new halo center is
determined, the sphere is regrown and the mass is determined.

All particles within $\rd$ are marked as members of a halo and skipped
when encountered in the loop over all particle densities. Particles
located just outside of a halo can be chosen as candidate
centers for other halos, but the iterative halo-centering procedure will wander into
the parent halo. Whenever two halos have centers that are within the
larger halo's $\rd$, the halo with the largest maximum circular
velocity, defined as the maximum of the circular velocity profile,
$V_c(r)=[GM(<r)/r]^{1/2}$, is taken to be the parent halo and the
other halo is discarded.

We allow halos to overlap. As long as the halo center does not reside
within $\rd$ of another halo, the algorithm identifies these objects
as distinct structures. This is in accord with X-ray or SZ
observations which would identify and count such objects as separate
systems. The overlapping volume may contain particles. Rather than
attempt to determine which halo each particle belongs to, or to divide
each particle between the halos, the mass is double-counted. No
solution to this problem is ideal, but we find that the total amount
of double-counted mass is only $\sim 0.75\%$ of all the mass located
within halos, with no dependence on halo mass. This parallels the
treatment of close pairs of clusters detected observationally. When
two X-ray clusters are found to have overlapping isophotal contours,
each system is treated individually and double counting of mass will
occur as well.

For each value of $\D$, the halo finder is run independently. Halo
mass functions are binned in bins of width 0.1 in $\log M$ with no
smoothing. Errors on each mass function are obtained by the jackknife
method; each simulation is divided into octants and the error on each
mass bin is obtained through the variance of the halo number counts as
each octant is removed from the full simulation volume
(cf. \citealt{zehavi_etal:05}, equation [6]). The jackknife errors
provide a robust estimate of both the cosmic variance, which dominates
at low masses, and the Poisson noise that dominates at high masses
(see \citealt{hu_kravtsov:03} for a the relative contributions of each
source of error as a function of halo mass). 

When fitting the data, we only use data points with error bars less
than 25\% to reduce noise in the fitting process. We note that
mass bins will be correlated (low-mass bins more so than high mass
ones). We do not calculate the full covariance matrix of each mass
function, so the $\chi^2$ values obtained from the fitting procedure
should be taken as a general guide of goodness of fit, but not as an
accurate statistical measure.  However, we note that the data from
multiple simulations in each mass range will be
uncorrelated, and the lack of a covariance matrix should not bias our
best-fit values for the mass function.

\begin{figure*}
\epsscale{1.0} 
\plotone{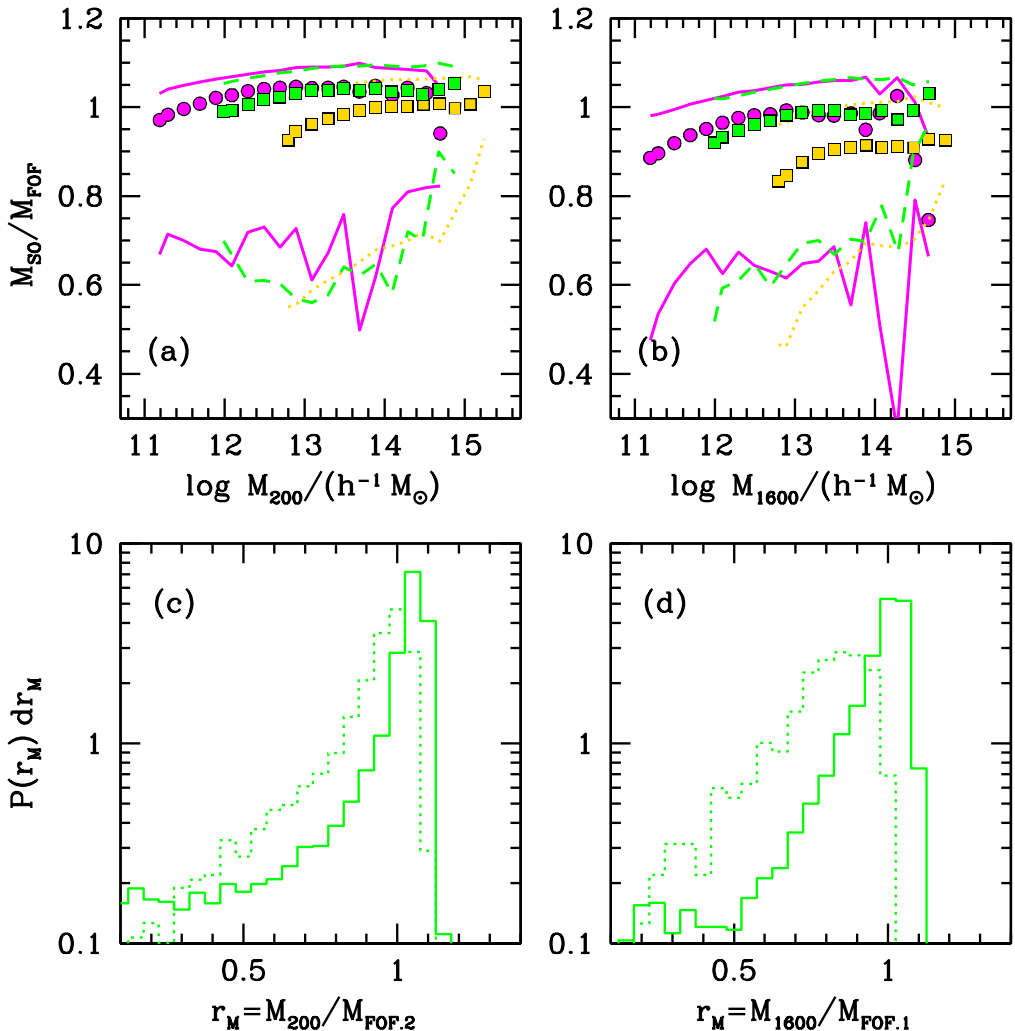}
\caption{ \label{fof_SO} Comparison between spherical overdensity
  masses and friends-of-friends masses for the same sample of objects
  from H384, L250, and L1000W. Panel (a) compares the masses of
  $\Delta=200$ halos to FOF halos with $l=0.2$. The symbols represent
  the median mass ratio, for objects binned by $M_{200}$. The curves
  show the upper 90\% and lower 10\% bounds of the distribution of
  mass ratios in each $M_{200}$ bin: solid for H384, dashed for L250,
  and dotted for L1000W. The asymmetry in the mass ratio distribution
  reflects the tendency of FOF to link objects together. Panel (b)
  compares $\Delta=1600$ halos to FOF objects with $l=0.1$. Panel (c)
  shows the distribution of mass ratios, $r_M = M_{200}/M_{\rm
    FOF.2}$, for halos $13\le \log\,M_{200}\le 14$ ({\it solid
    line}). The long tail of the distribution at $r_M<0.5$ indicates
  SO halos that are linked with other virialized objects in the FOF
  halo-finding process. The dotted line is the same distribution at
  $z=1.25$. Panel (d) shows the distribution of $r_M$ for the same
  mass range, for the $\D=1600$ and and FOF linking length
  $l=0.1$. Solid and dotted lines are $z=0$ and $z=1.25$,
  respectively. Both panels (c) and (d) show results for the L250
  run. }
\end{figure*}

\begin{figure}
\epsscale{1.25}
\plotone{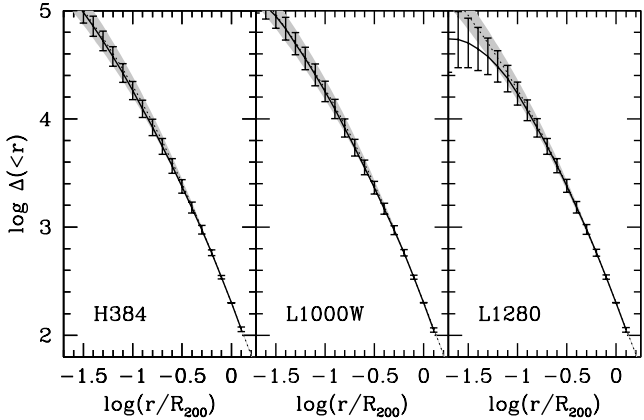}
\vspace{-2cm}
\caption{ \label{profiles} The halo density profiles are compared to
  analytic predictions for three different simulations. In each panel,
  the dotted curve represents the mean interior density given by an
  NFW profile with $c(M)$ from \citet{dolag_etal:04}. The shaded
  region is the expected scatter assuming $\sigma_{\log c}=0.12$. The
  solid curves with errorbars represent the numerical results. The
  left panel shows results from H384 for all halos $M>10^{14.5}$
  \hmsol. The center and right panels show results for halos
  $M>10^{15}$ \hmsol. The center and left panel demonstrate that halo
  profiles are well resolved in these simulations. The right panel,
shows significant deviations from the expected NFW profile
  at $r<0.1R_{200}$ in the lower-resolution L1280 simulation.  }
\end{figure}

\begin{figure}[t!]
\epsscale{1.15} 
\plotone{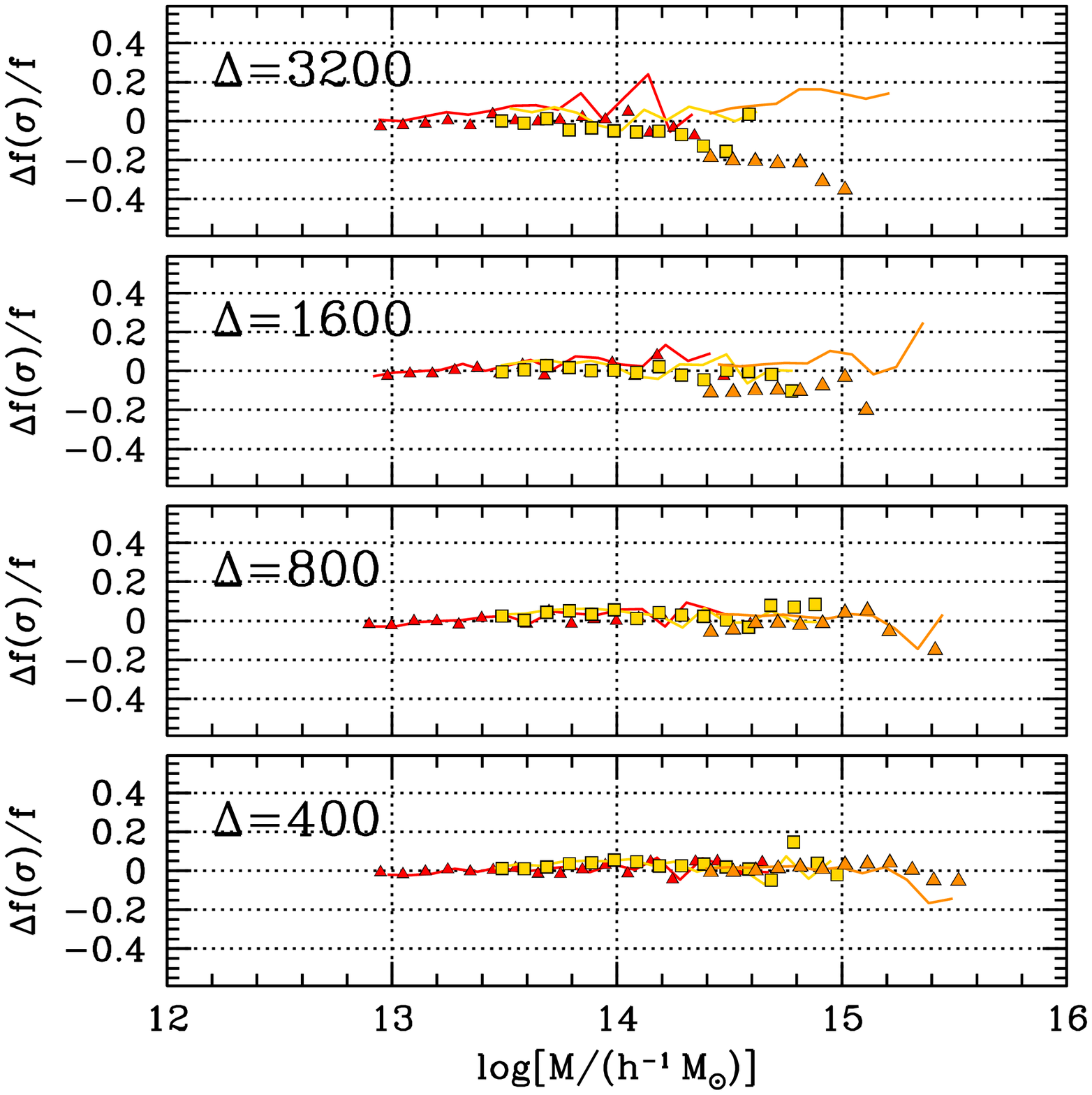}
\caption{ \label{resolution} Test of the resolution of the
  large-volume simulations, L500, L1000W, and one realization of
  L1280. In each panel, the mass functions are plotted as residuals
  with respect to the best-fit $\fsig$ function from Table 2. The
  symbols represent the mass functions measured directly from the
  simulations at each $\D$. The curves are mass functions inferred
  from the $\D=200$ halo catalog of each simulation, where the mass of
  each $\D=200$ halo is scaled to higher overdensities assuming an
  analytic NFW halo (including scatter in concentrations at fixed
  mass). For the two higher resolution simulations, the scaled and
  true mass function are in agreement. Due to insufficient resolution,
  the L1280 mass function falls below the scaled mass function at high
  $\D$. }
\end{figure}

\subsection{Comparison of FOF and SO halos}
\label{sec:fof_vs_so}

\cite{white:01, white:02} demonstrated that there is scatter between
the masses of halos identified with the FOF and SO halo definitions,
as well as an offset between the mean halo masses using the canonical
values of the linking length $\link=0.2$ in the FOF algorithm and
overdensity $\Delta=200$ in the SO approach. Figure \ref{fof_SO}
compares the masses of halos identified with these two definitions for
three different simulations. Halos in a simulation are first
identified with our SO approach, then the FOF finder is subsequently
run, beginning at the center of the SO halo. Figure \ref{fof_SO}a
compares $\D=200$ to $\link=0.2$.  The symbols represent the median
mass ratio $r_M = M_{200}/M_{\rm FOF.2}$ as a function of
$M_{200}$. The curves represent the upper and lower 90\% bounds on the
distribution of mass ratios. Although the median is near unity, the
scatter is large and highly asymmetric.

The asymmetry in the distribution is due to the FOF algorithm linking
two or more distinct objects in close proximity to each other. Because
we allow halos to overlap, FOF will treat these halos as a single
object. Due to the arbitrary shape of FOF halos, the algorithm also
links SO objects that do not overlap. The median mass ratio is also
sensitive to the number of particles per halo; FOF halos are known to
be biased toward higher masses at low particle number
(\citealt{warren_etal:06}).  The scatter between mass definitions is
not alleviated by making the linking length smaller. This is shown in
Figure \ref{fof_SO}b, in which the same results are shown for
$\D=1600$ and $\link=0.1$. The median is once again near unity, and
the scatter remains identical. We note also that there is an offset in
the median between simulations as well; the results from L1000W are
$\sim 5\%$ lower than the other simulations at $\link=0.2$ and $\sim
10\%$ lower at $\link=0.1$. This offset is not due to the change in
cosmology between the L1000W and the other simulations, therefore it
must be a result of the lower mass resolution.

We find that the curvature in the median mass ratio is alleviated when
adjusting the masses $M_{\rm FOF.2}$  by the Warren et.~al.\
correction formula, $(1-N_p^{-0.6})$, where $N_p$ is the number of
particles in a halo. However, the curvature is not entirely
ameliorated by this formula at $\link=0.1$, demonstrating that the
mass errors in FOF halos depend on the linking length. We find that
$(1-N_p^{-0.5})$ is sufficient to remove the FOF bias for
$\link=0.1$. Figures \ref{fof_SO}c and \ref{fof_SO}d show the
distribution of mass ratios for halos between $10^{13}$ and $10^{14}$
\hmsol\ for one of the simulations. The solid histograms present
results at $z=0$ and the dotted histograms is for $z=1.25$. Both the
$z=0$ histograms exhibit a large, constant tail to low ratios. At
higher redshift, the asymmetry of $P(r_M)$ becomes even stronger. 
The correlation between spherically-defined
masses and the FOF masses is thus broad and evolves with
redshift.

This has significant implications for comparisons with observational
cluster counts. Given that cluster observables correlate strongly with
the spherical overdensity masses, the large scatter between
$M_{\Delta}$ and $M_{\rm FOF}$ indicates that the FOF correlation will
be weaker. If one is to use a halo mass function calibrated against
halos and masses identified with the FOF algorithm, a significant
additional effort would be required to calibrate the scatter between
FOF masses and observables as a function of redshift, mass, and
cosmology.  In addition, this calibration will have to rely solely on
theoretical modeling, because the mass equivalent to the FOF cannot be
directly measured in observations.  The use of the halo abundance
predictions made with the spherical overdensity algorithm is therefore
strongly preferred.

\subsection{Accounting for effects of resolution \label{s.rescale}}

Defining the halo masses within a radius enclosing a given overdensity stipulates that
the halo mass is the integrated density profile within a fixed
radius. This means that the mass depends on the internal density
distribution of the halo, and is thus more susceptible to the
effects of resolution. The FOF masses,
on the other hand, are measured 
within a given isodensity surface, and are therefore less
sensitive to the internal mass distribution. 
For example, \cite{lukic_etal:07} demonstrate
that a reasonable FOF mass function can be obtained through a
low-resolution simulation with as little as 8 timesteps.
If the same simulation is performed twice with different resolutions, 
the same density peak in the lower resolution
simulation will have a shallower density profile and will 
in general have a different measured mass, $M_{\Delta}$.  The result is a
systematic artificial shift in the estimated halo mass function. This effect will
be larger for larger values of $\D$, as smaller radii that enclose
larger overdensities are more affected by resolution issues. 

To measure the SO mass function reliably at high $\D$, we 
test whether the halo density profiles are properly resolved in each of the
analyzed simulations at the overdensity in question. Figure
\ref{profiles} illustrates one of the resolution tests that
we performed. It compares the halo density profiles from simulations to the expected
profiles. For the latter we use the well-tested \cite{nfw:97} profile (hereafter NFW) with
the concentration for a given mass measured in high-resolution simulations by 
\cite{dolag_etal:04}\footnote{$c_{200}(M) = 9.59\times
  (M/10^{14})^{-0.102}$, normalized to the WMAP1 cosmology. When
  changing cosmology, we shift the normalization using the fractional
  change in concentration from the \cite{bullock_etal:01} model at
  $M=10^{13}$ \hmsol.} and a scatter in concentration of 0.12 in
$\log_{10} c$. In this figure we show examples of one HOT simulation
(H384), one ART simulation (L1000W), and one GADGET2 simulation
(L1280). The HOT and ART simulations have force resolutions of 14 and
30 \hkpc, respectively, which is well within the scale radius of a
typical cluster-sized halo. The results for both the mean profile and
its scatter are in excellent agreement with the NFW profile. The L1280
simulation has a force resolution of 120 \hkpc, and deviations from
the expected profile become clear at $r<0.1R_{200}$. These differences
will propagate into the estimate of the mass function if they are not
taken into account.

The results of comparisons similar to those shown in Figure
\ref{profiles} clearly identify which radii and which simulations
profiles are affected by resolution. These results can then be used to
determine the range of overdensities for which masses can be measured
reliably in a given simulation.  This is illustrated in
Figure~\ref{resolution}, which shows the mass functions from three
different simulations at four values of $\D$. The mass functions are
plotted relative to the best-fit mass functions at each $\D$, which
are described in more detail below in \S\ 3. At each overdensity we
compare the mass functions measured in simulations to mass functions
obtained by taking the individual halos found using the SO halo finder
with $\D=200$ and rescaling their masses assuming the NFW profile,
taking into account scatter in concentrations (see, e.g.,
\citealt{white:01, hu_kravtsov:03}). We use the concentration-mass
relation and scatter measured directly from our simulations (Tinker
et.~al., in preparation).  The figure shows that the measured and
re-scaled mass functions are in good agreement for $\D\leq 800$, where
the scaled-up mass function is $\sim 5\%$ higher than the true mass
function. This error is accrued from the halos located within
$R_{200}$, which can become separate halos for higher overdensities
and are not accounted for in the rescaling process.

At higher overdensities, the agreement is markedly worse, especially
for the lower-resolution L1280 boxes.  At $\D=1600$, the measured mass
function is underestimated by $\sim 10\%$, increasing to $\sim 20\%$
at $\D=3200$. Therefore, for this simulation we use the
directly-measured mass function only at $\D\le 600$, while at higher
$\D$ we calculate the mass function by mass re-scaling using halos
identified with an overdensity $\D=600$. A scaling baseline of
$\log(\Delta_{\rm high}/\Delta_{\rm low})\le 0.9$ accrues only
$\lesssim 2\%$ error in the amplitude of the mass function at these
masses. Thus the rescaled halo catalogs are reliable for calibrating
the halo mass function at high overdensity. This procedure is used to
measure high-$\D$ mass functions for L768 (for $\D>800$) and L1280
(for $\D>600$).

At $\D=200$ we choose a conservative minimum value of no less than 400 particles
per halo. Below this value resolution effects become
apparent, and simulations with differing mass resolutions begin to
diverge. This is readily seen in the SO mass functions analyzed in
\cite{jenkins_etal:01}. At higher $\D$, halos are probed at
significantly smaller radii, and the resolution requirements are more
stringent. Thus at higher $\D$ we increase the minimum number of
particles such that, at $\D=3200$, $N_{\rm min}$ is higher by a factor
of 4. Exact values for each overdensity are listed in Table 2.


\section{Halo Mass Function}

\subsection{ Fitting Formula and General Results }

\begin{figure}
\epsscale{1.2} 
\plotone{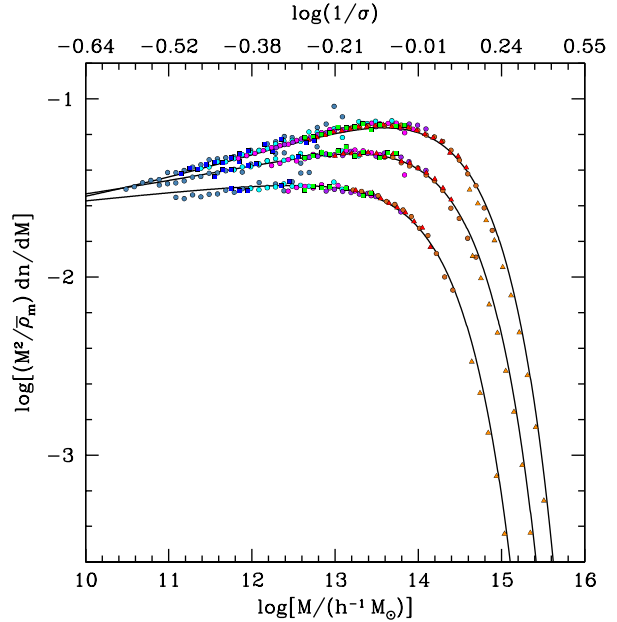}
\caption{ \label{dndM} The measured mass functions for all WMAP1
  simulations, plotted as $(M^2/\bar{\rho}_m)\,dn/dM$ against
  $\log\,M$. The solid curves are the best-fit functions from Table
  2. The three sets of points show results for $\D=200$, 800, and 3200
  (from top to bottom). To provide a rough scaling between $M$ and
  $\siginv$, the top axis of the plot shows $\siginv$ for this mass
  range for the WMAP1 cosmology. The slight offset between the L1280
  results and the solid curves is due to the slightly lower value of
  $\om=0.27$.}
\end{figure}

Although the number density of collapsed halos of a given mass depends
sensitively on the shape and amplitude of the power spectrum,
successful analytical ansatzes predict the halo abundance quite
accurately by using a universal function describing the mass fraction
of matter in peaks of a given height, $\nu\equiv
\delta_c/\sigma(M,z)$, in the linear density field smoothed at some
scale $R =(3M/4\pi\bar{\rho}_m)^{1/3}$
\citep{press_schechter:74,bond_etal91,sheth_tormen:99}.  Here,
$\delta_c\approx 1.69$ is a constant corresponding to the critical
linear overdensity for collapse and $\sigma(M,z)$ is the rms variance
of the linear density field smoothed on scale $R(M)$. The traditional
nonlinear mass scale $\mstar$ corresponds to $\sigma = \delta_c$. This
fact has motivated the search for accurate universal functions
describing simulation results by \cite{jenkins_etal:01},
\cite{white:02}, and \cite{warren_etal:06}.  Following these studies,
we choose the following functional form to describe halo abundance in
our simulations:
\begin{equation}
\label{e.dndsigma}
\frac{dn}{dM} = f(\sigma)\,\frac{\bar{\rho}_m}{M}\frac{d\ln \sigma^{-1}}{dM}.
\end{equation}
Here, the function $f(\sigma)$ is expected to be universal to the
changes in redshift and cosmology and is parameterized as
\begin{equation}
\label{e.fsig}
f(\sigma) = A\left[\left(\frac{\sigma}{b}\right)^{-a} + 1\right]e^{-c/\sigma^2}
\end{equation}
where
\begin{equation}
\sigma = \int P(k)\hat{W}(kR)k^2dk,
\end{equation}
and $P(k)$ is the linear matter power spectrum as a function of
wavenumber $k$, and $\hat{W}$ is the Fourier transform of the
real-space top-hat window function of radius $R$. It is convenient to
recall that the matter variance monotonically decreases with
increasing smoothing scale, thus higher $M$ corresponds to lower
$\sigma$.

\begin{figure*}
\epsscale{1.2}
\plotone{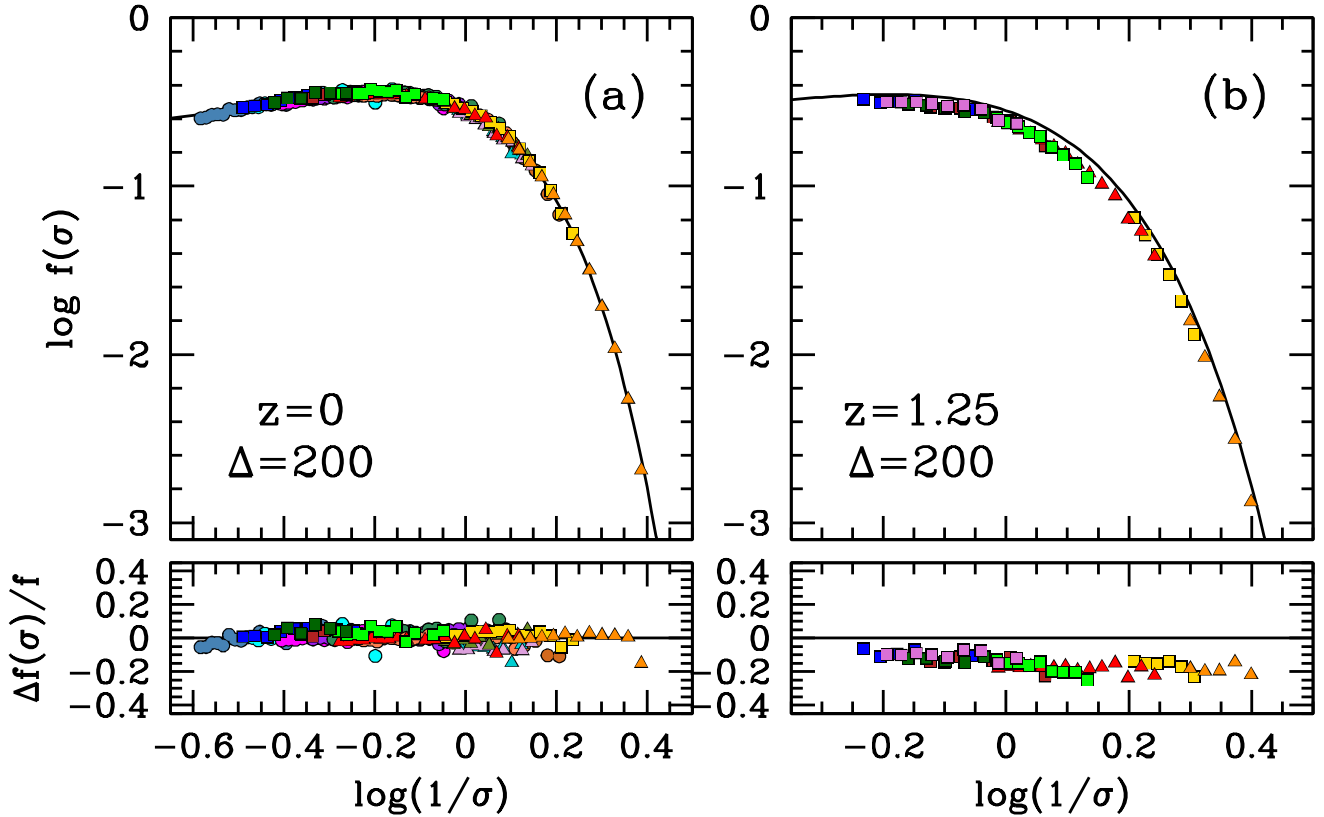}
\vspace{-5.5cm}
\caption {\label{signature} Panel (a): The measured $\fsig$ from all
  simulations in Table 1. Results are presented at $z=0$ and for
  $\D=200$. The solid line is the best fit function of equation
  (\ref{e.fsig}). The lower window shows the percentage residuals with
  respect to the fitting function. In the WMAP1 cosmology, the range
  on the data points on the $x$-axis is roughly $10^{10.5}$ \hmsol\ to
  $10^{15.5}$ \hmsol.  Panel (b): The measured $\fsig$ at $z=1.25$. We
  restrict results to simulations for which we have previous redshift
  outputs. In the WMAP1 cosmology, the range of data points on the
  $x$-axis is $10^{11}$ \hmsol\ to $10^{15}$ \hmsol. The solid line
  is the same as in panel (a), which was calibrated at $z=0$. The
  lower window shows that the $z=1.25$ mass function is offset by
  $\sim 20\%$ with respect to the results at $z=0$.}
\end{figure*}

The functional form~(\ref{e.fsig}) was used in \cite{warren_etal:06},
with minor algebraic difference, and is similar to the forms used by
\cite{sheth_tormen:99}\footnote{A convenient property of the Sheth \&
  Tormen mass function is that one recovers the mean matter density of
  the universe when integrating over all $\siginv$. Equation
  (\ref{e.fsig}) does not converge when integrating to $\siginv=0$. In
  Appendix C we present a modified fitting function that is properly
  normalized at all $\D$ but still produces accurate results at
  $z=0$.} and \cite{jenkins_etal:01}. Parameters $A$, $a$, $b$, and
$c$ are constants to be calibrated by simulations.  The parameter $A$
sets the overall amplitude of the mass function, while $a$ and $b$ set
the slope and amplitude of the low-mass power law, respectively. The
parameter $c$ determines the cutoff scale at which the abundance of
halos exponentially decreases.

 The best fit values of these parameters were determined by fitting
eq.~(\ref{e.fsig}) to all the $z=0$ simulations using $\chi^2$
minimization and are listed in Table 2 for each value of $\D$. For
$\Delta\ge 1600$, we fix the value of $A$ to be 0.26 without any loss
of accuracy\footnote{Although a four-parameter function is required to
accurately fit the data at low $\D$, at high overdensities the error
bars are sufficiently large that a degeneracy between $A$ and $a$
emerges, and the data can be fit with only three free parameters,
given a reasonable choice for $A$.}. This allows the other parameters
to vary monotonically with $\D$, allowing for smooth interpolation
between values of $\D$.

Figure~\ref{dndM} shows the mass function measured for three values of
$\D$ and the corresponding best fit analytic functions. We plot
$(M^2/\bar{\rho}_m)\,dn/dM$ rather than $dn/dM$ to reduce the dynamic
range of the $y$-axis, as $dn/dM$ values span nearly 14 orders of
magnitude. The figure shows that as $\Delta$ increases the halo masses
become systematically smaller. Thus from $\Delta=200$ to 3200, the
mass scale of the exponential cutoff reduces substantially. The shape
of the mass function is also altered; at $\Delta=200$ the logarithmic
slope at low masses is $\sim -1.85$, while at $\Delta=3200$ the slope
is nearly $-2$. This change in slope is due to two effects. First, the
change in halo mass accrued with changing the halo definition $\D$ is
not independent of mass. Because halo concentrations depend on mass,
$dM_{\D 1}$ does not equal $dM_{\D 2}$ for halos of two different masses. 

Second, a number of low-mass objects within $R_{200}$ of a larger halo
are ``exposed'' as distinct halos when halos are identified with
$\D=3200$. Although all halos contain substructure, these ``revealed''
subhalos will only impact overall abundance of objects at low mass,
$M\lesssim 10^{12}$ \hmsol, because the satellite fraction (the
fraction of all halos located within virial radii of larger halos)
decreases rapidly from $\approx 20\%$ to zero for $M>10^{12}$ \hmsol\
\cite[e.g.][]{kravtsov_etal:04}.  This trend can be understood using
average properties of subhalos in parent CDM halos. Subhalo
populations are approximately self-similar with only a weak trend with
mass \citep[e.g.,][]{moore_etal99,gao_etal04}, and the largest subhalo
typically has a mass of $\approx 5-10\%$ of the host mass.  Thus, at a
given mass $M$ only hosts with masses $>10M$ can produce significant
number of new halos when halo identification at higher $\D$ is
performed. At high masses, the number of such halos decreases
exponentially with mass, and therefore the contribution of such
``exposed'' halos becomes small.

\begin{figure*}
\epsscale{1.0} 
\vspace{-1.8cm}
\plotone{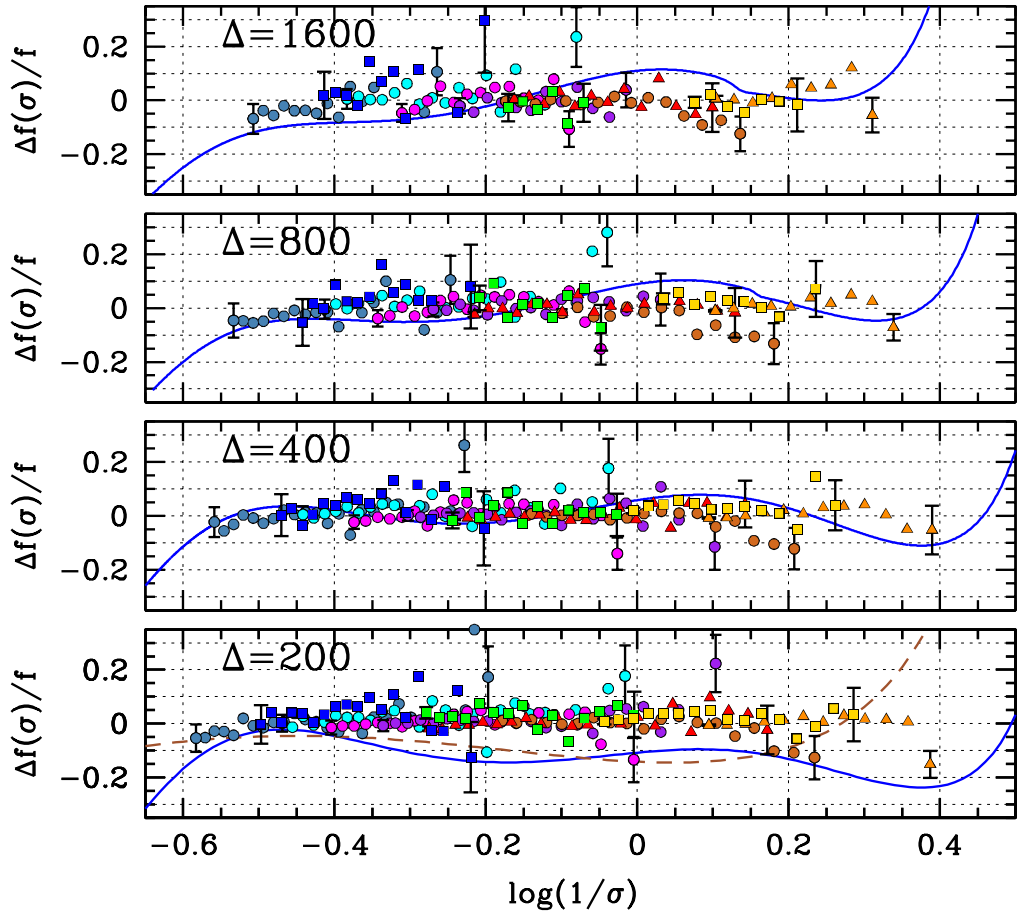}
\caption{ \label{diff_wmap1} Residuals of the measured mass functions
  with respect to the best fit analytic mass functions from Table 2
  for all WMAP1 simulations at $z=0$. Error bars are shown for the
  first and last point for each simulation, and only points with less
  than 10\% error bars are plotted, with the exception of L80, for
  which 15\% is the maximum.  For $\D=200$, the blue curve represents
  the \cite{jenkins_etal:01} SO180 mass function (scaling up to
  $\D=200$ yields indistinguishable results). The red dashed curve
  represents the \cite{sheth_tormen:99} mass function. For $\D=400$, the blue
  curve represents the \cite{jenkins_etal:01} SO324 (scaled up to
  $\D=400$). For $\D=1600$ and $\D=800$, the solid curve represents
  the Jenkins SO(324) mass function scaled up analytically assuming
  NFW profiles. }
\end{figure*}

\begin{figure*}
\epsscale{1.0} 
\plotone{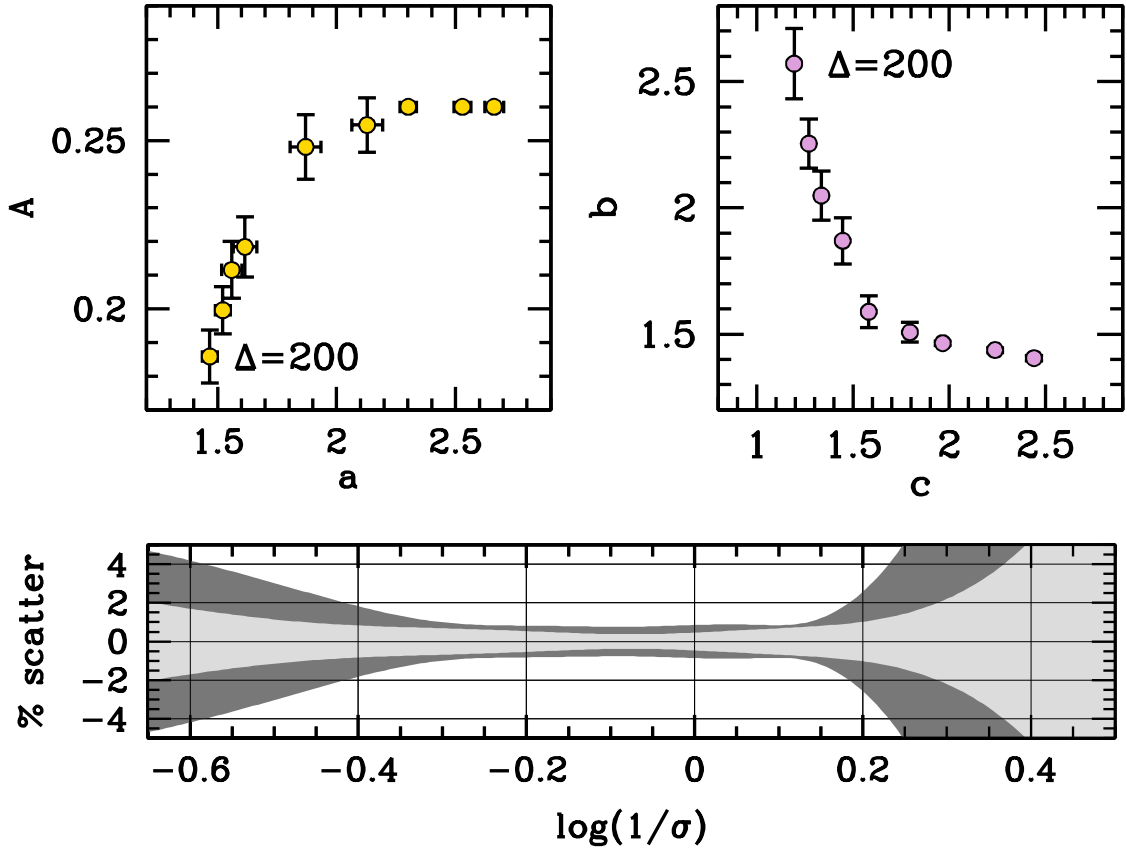}
\vspace{-2.9cm}
\caption{ \label{params} The trajectories of the best-fit parameters
  of $\fsig$ from Table 2. In each panel, the order of the points is
  from low-$\D$ to high-$\D$ (left to right). Error bars represent
  1-$\sigma$ variance of parameters from the MCMC chain. In panel (a),
  the normalization $A$ is plotted against power-law slope $a$. In
  panel (b), the power-law amplitude $b$ is plotted against the
  cutoff scale $c$. The lower panel shows the rms scatter of mass functions
  from 100 bootstrap samples, creating by sampling the simulation
  list. Light gray is for $\D=200$, while dark gray is for
  $\D=1600$. }
\end{figure*}


\begin{deluxetable*}{cccccccccc}
\tablecolumns{10} 
\tablewidth{31pc} 
\tablecaption{Mass Function Parameters for $\fsig$ at $z=0$}
\tablehead{\colhead{$\D$} & \colhead{$A$} & \colhead{$a$} & \colhead{$b$} &\colhead {$c$} & \colhead{$\xdof$ (ALL)} & \colhead{$N_{\rm min}$} & \colhead{$\xdof$ (WMAP1)} & \colhead{$\xdof$ (WMAP3)} & \colhead{$\xdof$ (WMAP3-fit)} 
}
\startdata

200 & 0.186 & 1.47 & 2.57 & 1.19 & 1.15 & 400 & 1.07 & 1.66 & 1.62\\
300 & 0.200 & 1.52 & 2.25 & 1.27 & 1.17 & 400 & 1.08 & 1.65 & 1.60 \\
400 & 0.212 & 1.56 & 2.05 & 1.34 & 1.05 & 600 & 0.96 & 1.49 & 1.37 \\
600 & 0.218 & 1.61 & 1.87 & 1.45 & 1.06 & 600 & 0.99 & 1.55 & 1.28 \\
800 & 0.248 & 1.87 & 1.59 & 1.58 & 1.10 & 1000 & 1.07 & 1.36 & 1.14\\
1200 & 0.255 & 2.13 & 1.51 & 1.80 & 1.00 & 1000 & 0.97 & 1.22 & 1.16\\
1600 & 0.260 & 2.30 & 1.46 & 1.97 & 1.07 & 1600 & 1.03 & 1.34 & 1.25\\
2400 & 0.260 & 2.53 & 1.44 & 2.24 & 1.11 & 1600 & 1.07 & 1.50 & 1.26\\
3200 & 0.260 & 2.66 & 1.41 & 2.44 & 1.14 & 1600 & 1.09 & 1.61 & 1.33\\

\enddata \tablecomments{$N_{\rm min}$ is the minimum number of
  particles per halo used in the fit. Fits are for simulations at
  $z=0$. The WMAP1 and WMAP3 $\xdof$ values are with respect to the
  WMAP1 and WMAP3 simulations, respectively, but using the best-fit
  parameters. The WMAP3-fit $\xdof$ values are independent fits using
  only the WMAP3 simulations}

\end{deluxetable*}

\begin{figure*}
\epsscale{1.0} 
\vspace{-1.8cm}
\plotone{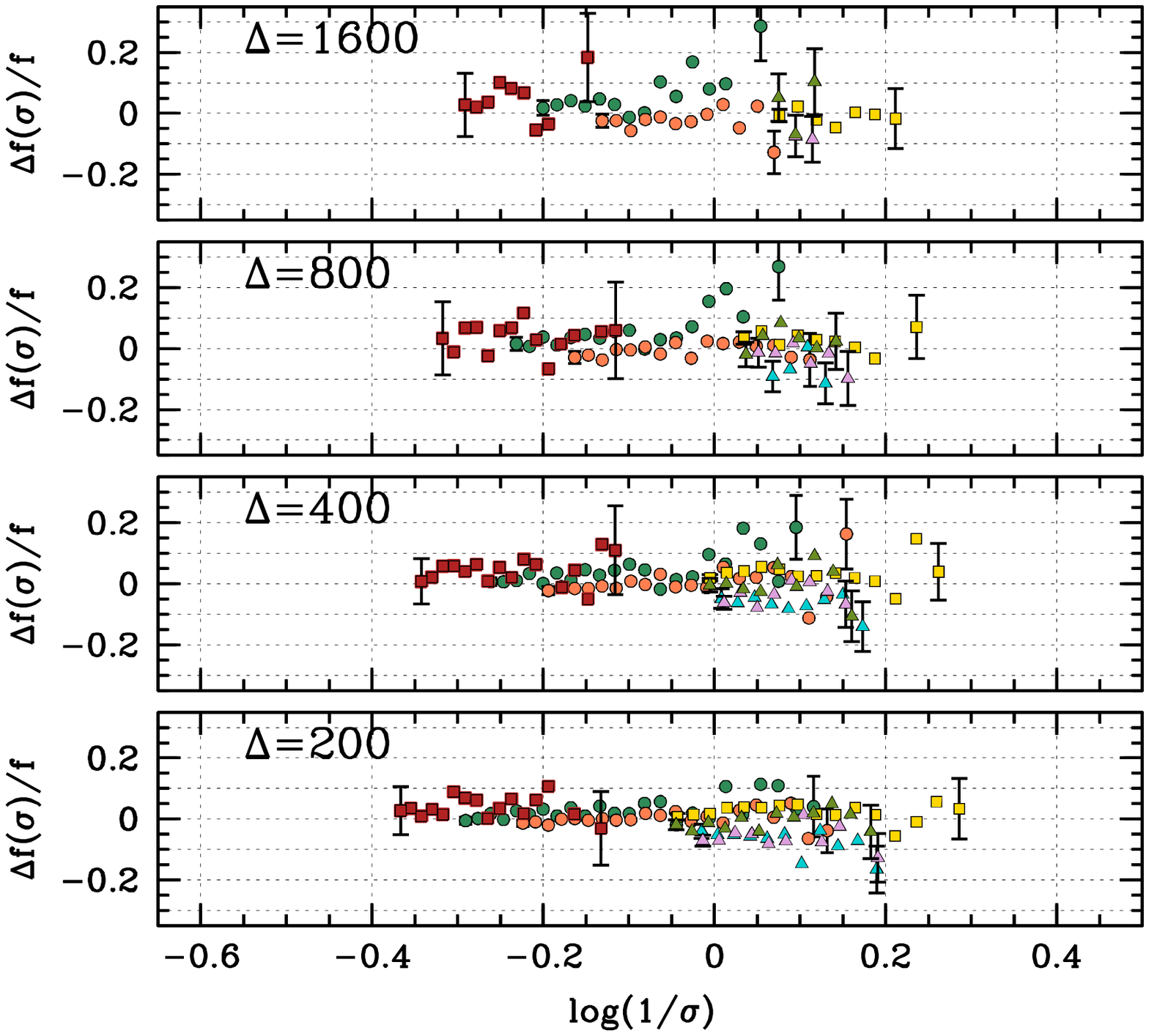}
\caption{ \label{diff_wmap3} Residuals of the measured mass functions
  with respect to the best fit analytic mass function from Table 2 for
  all WMAP3 simulations at $z=0$.  Error bars are shown for the first
  and last point for each simulation, and only points with less than
  10\% error bars are plotted, with the exception of L80W, for which
  15\% is the maximum. }
\end{figure*}

Figure \ref{signature}a shows the function $\fsig$ measured for all
simulations in Table 1 at $z=0$ with $\D=200$. The solid curve is
equation (\ref{e.fsig}) using the best-fit parameters from Table
2. The residuals with respect to this fit demonstrate the high
accuracy of our numerical results and the consistency of different
codes, mass resolutions, and cosmologies. Figure \ref{signature}b
shows $\fsig$ at $z=1.25$ for a subset of simulations for which higher
redshift outputs are available. The solid curve represents the results
from $z=0$. At this redshift, the results at $\sim 20\%$ below the
$z=0$ results, nearly independent of $\siginv$. This demonstrates that
the mass function is {\it not} universal in redshift, or for
correspondingly large changes in cosmology,\footnote{Note that we can
interpret higher redshift outputs of a given simulation as the $z=0$ epoch
of a simulation with different cosmological parameters corresponding
to $\Omega_m(z)$ and other parameters at the redshift in question.}
at this level of accuracy. We address evolution of $\fsig$ with $z$ in
\S 3.3 below.

\subsection{Results as a function of $\D$}

The best-fit parameters of equation (\ref{e.fsig}) resulting from fits to 
{\it all} simulations for 9 values of
overdensity are listed in Table 2. Figure \ref{diff_wmap1} shows  the
residuals of individual WMAP1 simulations with respect to global 
fits at different $\D$. We include L1000W in these
plots to show consistency between cosmologies at cluster masses. For
the fifty realizations of L1280, we plot the mean $\fsig$ and the
error in the mean. Each panel shows the fractional residuals of the
measured mass functions with respect to the best-fit $\fsig$ for four
values of $\D$. To avoid crowding, error bars are plotted for the
maximum and minimum mass scale for every simulation; the latter is representative
of the cosmic variance given the finite simulation volume, while the former is
dominated by Poisson noise. We list formal values of $\xdof$  for our diagonal
error bars in Table 2. The values in column 6 are for all $z=0$
simulations, while the value in column 8 is the $\xdof$ for the
same parameters but with respect to the WMAP1 simulations only. Not
surprisingly, the $\xdof$ values reduce slightly when comparing the
best-fit $\fsig$ to the WMAP1 simulations only, which comprise 
a majority of the simulations and therefore drive the fitting
results.

The solid blue curve in the $\D=200$ panel represents the fitting
function of \cite{jenkins_etal:01} calibrated on their set of
$\tau$CDM simulations (their equation B3), using $\D=180$ (rescaling
this equation to 200 yields nearly indistinguishable results). At
$M\gtrsim 10^{12}$ \hmsol, the Jenkins result is $10$--$15\%$ below 
our results. The \cite{sheth_tormen:99} function is similarly offset from our
results.  In the $\D=400$ panel, the blue curve shows the Jenkins
et.~al. fitting function calibrated to $\D=324$ on their set of \lcdm\
simulations (essentially the WMAP1 cosmology). For this comparison the
Jenkins formula has been rescaled to $\D=400$ using the same halo
rescaling techniques discussed in \S 2.3 and in
\cite{hu_kravtsov:03}. The Jenkins SO(324) function (their Equation
B4) is in good agreement with our results for $M<10^{13}$ \hmsol, while
at higher masses there are variations of $\pm 5$--$10\%$.

The solid curves in the $\D=800$ and 1600 panels are the Jenkins
SO(324) result scaled up to those overdensities. At $\siginv>1$, the
residuals increase, while at lower masses the rescaled $\fsig$
underestimates the numerical results by $5-10\%$. Both of these
effects are due to subhalos becoming exposed when halos are
identified using higher overdensity. If a high-mass halo contains a large
subhalo, the rescaling procedure will overestimate the
mass of that object at higher $\D$. At low masses, the rescaling
procedure does not account for the revealed substructures. The change
in mass from $\D=200$ to $\D=1600$ is $\sim 50\%$ at $10^{14}$
\hmsol. If subhalos are distributed within parent halos in a similar fashion to
the dark matter, then the rescaling procedure should underestimate
the mass function by $\sim 0.5\times 0.2=0.1$ (where 0.2 is the
subhalo fraction for low-mass halos from \citealt{kravtsov_etal:04}).

Figure \ref{params} shows that the best fit parameters of $\fsig$ vary with
$\D$ smoothly. This means that interpolating
between these best-fit parameters can be expected to yield accurate mass function parameters at any
desired overdensity. In Appendix B we show examples of the
interpolated mass functions, as well as fitting function for the
$\fsig$ parameters as a function of $\D$. The error bars are
1$\sigma$ and are obtained by marginalizing over all other parameters. The
errors on the amplitude $A$ are $\sim 3-4\%$, but this parameter is
highly correlated with $b$ and the true scatter about the best-fit
$\fsig$ is $\lesssim 1\%$ at most masses.

The lower panel in Figure \ref{params} shows the rms scatter in our
constraints on $\fsig$. The scatter was calculated by bootstrap
resampling of the simulation set and repeating the fitting process on
100 bootstrap samples.\footnote{Because the 50 realizations of L1280
outnumber all the rest of the simulations (which only number 17), we
create bootstrap samples by first sampling from the list of L1280
realizations, then sampling from the rest of the simulation set. This
guarantees a fair sampling of the range of $\siginv$ probed by the
simulations. If we do not do this, many bootstrap samples will only
contain mass function with results above $M\gtrsim 2\times 10^{14}$
\hmsol, which would artificially inflate the size of the low-mass
errors.}  The shaded area is the variance of the bootstrap fits. The
light gray region represents results for $\D=200$, while the dark gray
region represents $\D=1600$. Between $\siginv=0.63$ and $\siginv=1.6$
the scatter is less than $1\%$ ($M=10^{11.5}$ \hmsol\ and $10^{15}$
\hmsol\ for the WMAP1 cosmology). Outside this mass range the results
diverge due to lack of coverage by the simulations.  Because the WMAP1
simulations dominate by number, these constraints should be formally regarded
as the accuracy of the fit for the WMAP1 cosmology.

Figure \ref{diff_wmap3} compares the calibrated mass functions from
Table 2 with the measured mass functions from the WMAP3 simulations
(i.e., the last seven entries in Table 1). Column 9 of Table 2
contains values of $\xdof$ for the WMAP3 simulations only. The $\xdof$
are somewhat larger than for the WMAP1 runs at all overdensities, even
though the WMAP3 residuals do not seem to be systematically offset
from the global $\fsig$ fits. We test this statistically by refitting
for the parameters of equation (\ref{e.fsig}) using {\it only} the
WMAP3 simulations. The $\xdof$ values are listed in column 10 of Table
2. For each $\D$, the $\chi^2$ of the fit is only reduced
marginally. In the mass range covered by our simulations, the
difference between the global $\fsig$ functions and those derived from
the WMAP3 simulations differ by $\lesssim 2\%$, but with a $\sim 4\%$
uncertainty in the normalization of the WMAP3-only fitting function,
derived from the bootstrap method described above. Thus we conclude
that the higher $\chi^2$ values are not due to a systematic change in
$\fsig$ due to variations in cosmology, but rather scatter in the
simulations themselves at the $\sim 5\%$ level, excluding obvious
outliers where Poisson noise dominates.

\begin{figure}
\epsscale{1.2} 
\plotone{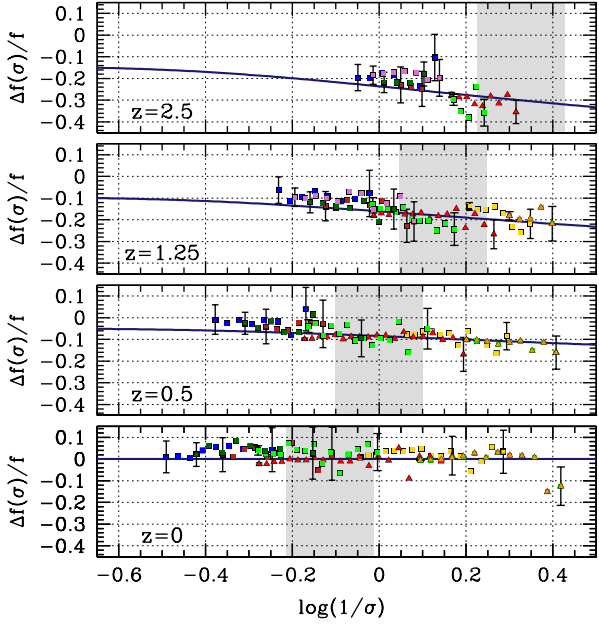}
\caption{ \label{redshift_200} Redshift evolution of the $\D=200$ mass
  function. Each panel shows the residuals of the $z=0$ mass function
  with respect to the measured mass functions at $z=0$, 0.5, 1.25, and
  2.5. Note that the simulation set used here is a combination of
  WMAP1 and WMAP3 boxes. Error bars are shown for the first and last
  points for each simulation, and only points with $<10\%$ are shown,
  with the exception of the L80 and L80W, for each 15\% is the
  limit. The shaded region brackets $10^{13}$ \hmsol\ to $10^{14}$
  \hmsol. The solid curves represent the $z=0$ mass function modified
  by equations (\ref{e.fsigz_A})---(\ref{e.fsigz_alpha}). }
\end{figure}

\begin{figure}
\epsscale{1.2} 
\plotone{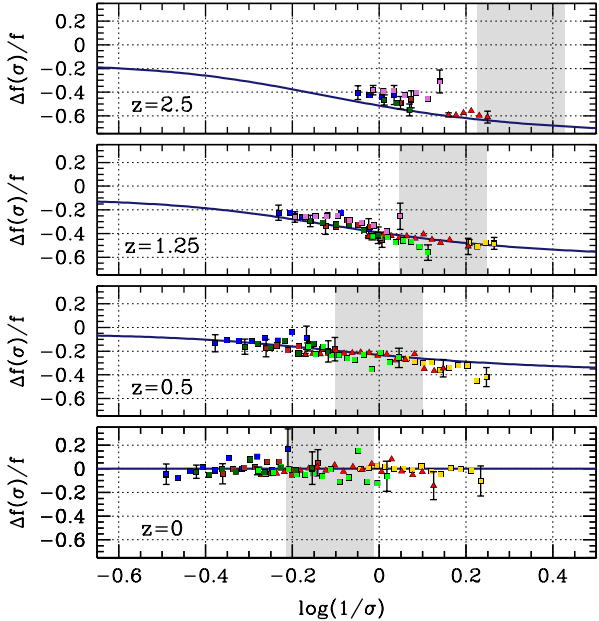}
\caption{ \label{redshift_1600} Redshift evolution of the $\D=1600$
  mass function. Each panel shows the residuals of the $z=0$ mass
  function with respect to the measured mass functions at $z=0$, 0.5,
  1.25, and 2.5. Note that the simulation set used here is a
  combination of WMAP1 and WMAP3 boxes. Results are plotted down to
  halos with 400 particles, as opposed to the limit of 1600 used in
  fitting $\fsig$. All points with errors $<15\%$ are plotted. The
  shaded region brackets $10^{13}$ \hmsol\ to $10^{14}$ \hmsol. The
  solid curves represent the $z=0$ mass function modified by equations
  (\ref{e.fsigz_A})---(\ref{e.fsigz_alpha}). }
\end{figure}

\begin{figure*}
\epsscale{0.85} 
\plotone{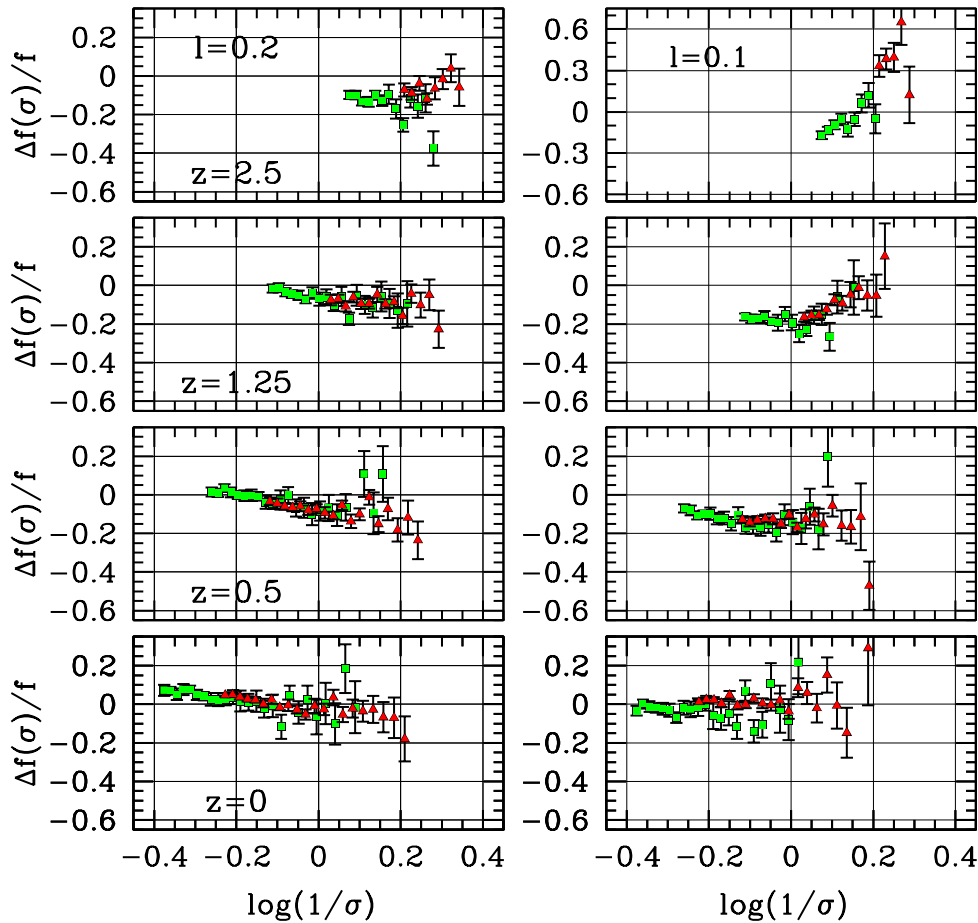}
\caption{ \label{fof} Evolution of the FOF mass function for linking
  lengths of $l=0.2$ (left panels) and $l=0.1$ (right panels). The
  simulations used are L500 and L250. The L500 simulation has been
  downsampled to 1/8 the original particle number. For $l=0.2$, the
  residuals are plotted with respect to the \cite{warren_etal:06}
  function. Mass functions are plotted down to 100 particles per halo
  but have not been corrected for discreteness effects (i.e., equation
  2 in Warren et.~al.). For $l=0.1$, the residuals are plotted with
  respect to the $\D=1600$ mass function from Table 1. Note the larger
  range of the $y$-axis at $z=2.5$ for $\link=0.1$. The FOF mass
  function evolves less than the SO mass function, but this largely a
  numerical effect due to increased linking of distinct halos. }
\end{figure*}

\subsection{Redshift Evolution}

Figure \ref{redshift_200} shows the evolution of the $\D=200$ mass
function for four different redshifts from $z=0$ to 2.5. Results are
plotted for the subset of simulations for which we have previous
redshift outputs. When modeled as pure amplitude evolution, the mass
function evolves as $(1+z)^{-0.26}$. However, it is also clear that
the shape of $\fsig$ is evolving with redshift such that the
amplitude at $\siginv>1$ decreases at a higher rate. This is more
evident in Figure \ref{redshift_1600}, in which $\fsig$ at $\D=1600$
is shown for the same redshifts. As $\D$ increases, both the evolution
in the amplitude and shape of $\fsig$ become stronger.

In Figures \ref{redshift_200} and \ref{redshift_1600}, the solid
curves show a model in which the first three parameters of $\fsig$ are
allowed to vary as a power law of $1+z$;

\begin{equation}
\label{e.fsigz_A}
A(z) = A_0\,\left(1+z\right)^{-0.14},
\end{equation}
\begin{equation}
\label{e.fsigz_a}
a(z) = a_0\,\left(1+z\right)^{-0.06} ,
\end{equation}
\begin{equation}
\label{e.fsigz_b}
b(z) = b_0\,\left(1+z\right)^{-\alpha}, 
\end{equation}
\begin{equation}
\label{e.fsigz_alpha}
\log \alpha(\D) = -\left(\frac{0.75}{\log(\D/75)}\right)^{1.2},
\end{equation}

\noindent where subscript `0' indicates the value obtained at $z=0$ in
Table 2. Modulation of $A$ controls the overall amplitude of $\fsig$,
while $a$ controls the tilt, and $b$ sets the mass scale at which the
power law in $\fsig$ becomes significant. Modifying $b$ results in a
shift between the amplitudes at low and high $\siginv$, thus it
encapsulates the changes in $\fsig$ with $\D$ seen in Figures
\ref{redshift_200} and \ref{redshift_1600}. Although the redshift
scaling introduced here matches the results at $z\le 2.5$ accurately,
residuals of $\gtrsim 5\%$ emerge at $z=2.5$. It is possible that the
evolution between $z=1.25$ and 2.5 is slowing down.  Because the
numerical results at $z=2.5$ are quite noisy and cover only a small
range in $\siginv$, our results at this value of $z$ and extrapolation
to higher redshifts must be checked with other
simulations. Extrapolating equation
(\ref{e.fsigz_A})-(\ref{e.fsigz_alpha}) to $z=10$ produces an $\fsig$
that is reduced by $\sim 50\%$ with respect to $z=0$. This seems
unlikely given current studies but needs to be checked with a
consistent halo finding algorithm.

\cite{reed_etal:07} parameterize the redshift-dependent mass function
in terms of both $\sigma$ and the effective spectral index of the
linear power spectrum, $n_{\rm eff}$. These authors use this
parameterization to model the mass function at $z>10$, where
differences in the slope of $n_{\rm eff}$ from $z=0$ are large. This
approach is ill suited for modeling the evolution at $z<3$, where
there is very little change in the effective spectral index.

It is interesting to note that the evolution in the exponential cutoff
scale is minimal. Any evolution in this mass scale would yield
quantitatively different residuals than those seen in Figure
\ref{redshift_200} and \ref{redshift_1600}. Namely, the residuals would show
pronounced curvature at $\siginv>1$. Our results show that the
dominant effect is a shift in the normalization in the mass function
rather than the cutoff mass scale. Thus our results are not consistent with
$\fsig$ being universal as a function of {\it virial} overdensity
because $\D_{\rm vir}$ evolves with redshift. Nor are our results
consistent with the mass function being universal at a fixed
overdensity with respect to the critical density (rather than defining
$\D$ with respect to the background, as we do here).

The \cite{jenkins_etal:01} study reports no detected evolution of the
FOF or SO mass functions with redshift. More recent results quantify
the evolution of the FOF at high redshift, $z\gtrsim 10$, to be
$5-10\%$ (\citealt{lukic_etal:07, reed_etal:07, cohn_white:07}).
However, friends-of-friends identified halos may have a different
response to changes in the redshift evolution of halo
profiles. Merging rates vary with redshift, and this may be reflected
in the FOF tendency to bridge distinct structures. Figure \ref{fof}
shows the redshift evolution for friends-of-friends selected halos in
the L500 and L250 boxes. The panels in the left column show the
results for halos identified with a linking length of 0.2. Residuals
are calculated with respect to the \cite{warren_etal:06} fitting
formula with their best fit parameters, plotted down to halos
containing 100 particles. The friends-of-friends masses have not been
corrected for any systematic errors (equation [2] in
\citealt{warren_etal:06}), resulting in the slight negative slope to
the residuals at low masses. The mass function shows some redshift
evolution, but only of order $\sim 10\%$ at z=1.25, or roughly half
that in Figure \ref{redshift_200}.

The right column shows the results for halos identified with a linking
length of $l=0.1$. The smaller linking length identifies halos with
higher overdensities. The residuals are with respect to $\fsig$
for $\D=1600$. For this linking length, the redshift evolution is
stronger than for $\link=0.2$, and the shape of the FOF mass function
changes dramatically. As a whole, these results indicate that the mass function
is also non-universal for FOF halos, with the degree of non-universality
depending on the linking length used. 

These results are in general agreement with those of other recent
studies that considered evolution of the mass function for FOF halos,
although the overall picture of how the mass function evolves with
redshift is not yet clear. The simulation results of
\cite{lukic_etal:07} exhibited $\sim -5\%$ residuals with respect to
the $z=0$ Warren et.~al.\ mass function as $z=5$, but with a monotonic
trend of rising residuals with increasing redshift. The FOF mass
function in the Millennium Simulation, shows roughly $20\%$
evolution from $z=0$ to 10 (\citealt{reed_etal:07}). Finally,
\cite{fakhouri_ma:07} recently showed that the Millennium Simulation
FOF mass function, once corrected for spurious FOF linking between
halos, evolves by $\sim 20\%$ from $z=0$ to 1. This is consistent
with our findings, but note that the volume of the Millennium
simulation and statistics at large masses is substantially worse than
in our set of simulations.

\begin{figure}
\epsscale{1.2} 
\plotone{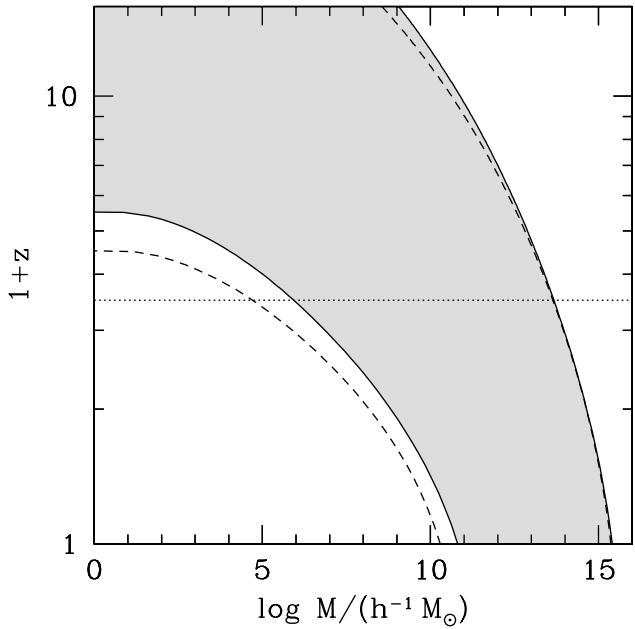}
\caption{ \label{mass_range} Halo mass range corresponding to the
  range of $\siginv$ on which $\fsig$ is calibrated. The shaded region
  bounded by the solid curves shows how this mass range evolves with
  redshift for the WMAP1 cosmology. The dashed curves show the upper
  and lower mass limits for the WMAP3 cosmology of the L80W
  simulation. The dotted line indicates the maximum redshift output of
  our simulation set. }
\end{figure}


\section{Summary and Discussion}

We have presented a new fitting function for halo abundances and their
evolution in the $\Lambda$CDM cosmology. The fitting function can be
used to predict halo mass functions for spherical aperture masses
defined with an arbitrary overdensity over a wide range of values.
For the WMAP1 cosmology our results are accurate at the percent level
in the mass range relevant for cluster cosmology. For the WMAP3
cosmology our results are accurate to $\lesssim 5\%$. One of our main
results is that the mass function is non-universal, and varies in a
systematic way with redshift in the interval $z=[0,2.5]$, with the
abundance of halos at a given $\siginv$ monotonically decreasing with
increasing $z$.

We have parameterized redshift evolution of $\fsig$ as a simple scaling
of the $z=0$ fitting parameters with $(1+z)^{\alpha}$. We note that if this
evolution is driven by changes in $\Omega_m$ with $z$, it may be more
robust to model $f(\sigma,z)$ as a function of the growth rate rather
than $1+z$. Our simulation set does not probe a large enough
cosmological parameter space to detect differences due to different
growth factors. However, this will become important when investigating
how the mass function evolves in dark energy cosmologies, in which the
primary change in structure formation is a different growth function
of perturbations.

Our finding of evolving, non-universal $\fsig$ is quantitatively
different from the results of previous analyses that use the
friend-of-friends method for halo identification, which generally show
weaker evolution and greater degree of universality of the function
$\fsig$. We argue that the likely explanation for this difference is
greater sensitivity of the SO defined mass to the redshift evolution
of halo concentrations. As discussed previously, SO masses are the
integrated halo profiles within a specified radius and lower halo
concentrations result in lower masses at fixed abundance (or,
conversely, fewer halos at fixed mass). The fact that the high-mass
end of the mass function (where concentrations at $z=0$ are lower and
the mass within $R_{200}/c_{200}$ is a significant fraction of the
total mass) evolves somewhat faster than the low-mass end, argues that
evolution of concentrations plays a significant role in the evolution
of $\fsig$.

The evolution of halo concentrations is mostly driven by the change in
$\Omega_m$ with redshift. This implies that $\fsig$ in cosmologies
with substantially different matter densities at $z=0$ will be
systematically different from the one we find here (perhaps closer to
our $z>1$ results).  There are indications that this is indeed the
case.  The H384$\Omega$ simulation, with $\Omega_m=0.2$, is above
$\fsig$ by $\sim 5\%$ at $z=0$. The \cite{jenkins_etal:01} fitting
function for $\D=180$ was calibrated on simulations with $\Omega_m=1$,
producing a fit $\sim 15\%$ below our results at the same
overdensity. The Jenkins SO(180) mass function is close to our
$\D=200$ results at $z=1.25$, where $\om$ is approaching unity.

The lower evolution of the FOF mass function with redshift can be
understood from Figure \ref{fof_SO}. The distribution of mass ratios
between FOF and SO halos changes between $z=0$ and $z=1.25$. The
median mass ratio, $M_{\rm SO}/M_{\rm FOF}$, decreases while the
scatter increases at higher $z$ due to more linking of {\it distinct}
objects.  The number of distinct objects at a fixed $\siginv$
decreases, but the higher incidence of linking offsets this
effect. Thus the weaker evolution of the FOF mass function is due to
this linking of separate collapsed halos and is largely
artificial. The better universality of $\fsig$ may still seem like an
advantage of the FOF mass function. However, as we discussed in this
paper, the large and redshift-dependent scatter between SO and FOF
masses implies similarly large and redshift-dependent scatter between
FOF masses and cluster observables. This makes robust interpretation
of observed cluster counts in terms of the FOF halo counts
problematic.

Our fitting function is calibrated over the range $0.25 \lesssim
\siginv \lesssim 2.5$, which at $z=0$ spans a range of halo masses
roughly $10^{10.5} \lesssim M \lesssim 10^{15.5}$ \hmsol, depending on
the specific choice of cosmology. In Figure \ref{mass_range} we show
how this mass range evolves with redshift. By $z=3$, the lower mass
limit is $\sim 10^5$ \hmsol. At this redshift, our fitting function is
in agreement with the numerical results of \cite{colin_etal:04}, which
probe the mass range $10^5\le M\le 10^9$ \hmsol. At higher redshifts,
$\siginv$ is a slowly varying function of mass, making the lower mass
limit evolve rapidly. Because our calibration of the redshift
dependence of the mass function parameters extends only to $z=2.5$, we
caution against extrapolation of equations
(\ref{e.fsigz_A})---(\ref{e.fsigz_alpha}) to significantly higher
redshifts. As noted above, $\fsig$ is evolving less rapidly from
$1.25<z<2.5$ than from $0<z<1.25$. Thus using the $z=2.5$ $\fsig$
should yield a mass function with reasonable accuracy at higher $z$,
but must be verified with additional simulations.

The range of cosmologies probed here is narrow given the volume of
parameter space, but it is wider than the allowed range given recent
results from CMB in combination with other large-scale measures
(\citealt{komatsu_etal:08}). For general use that does not require 5\%
accuracy, extending our results somewhat outside this range will
produce reasonable results. It is unlikely that variations in the
shape and amplitude of the power spectrum will yield significantly
different forms of $\fsig$. As discussed above, however, large
variations in $\om$ at $z=0$ (ie, $\om=0.1$ or $\om=1$), are not
likely to be fit by our $z=0$ mass function within our $5\%$
accuracy. Models with a higher matter density at $z=0$ can be
approximated by using our calibrated $\fsig$ at the redshift for which
$\om(z)$ is equal to the chosen value.

The next step in the theoretical calibration of the mass function for
precision cosmology should include careful examination of subtle
dependencies of mass function on cosmological parameters (especially
on the dark energy equation of state), effects of neutrinos with
non-zero mass, effects of non-gaussianity
\citep{grossi_etal07,dalal_etal08}, etc.  Last, but not least, we need
to understand the effects of baryonic physics on the mass distribution
of halos and related effects on the mass function, which can be quite
significant \citep{rudd_etal08}. The results of \cite{zentner_etal:07}
indicate that the main baryonic effects can be encapsulated in a
simple change of halo concentrations, which would result in a uniform
shift of $M_{\Delta}$ and a uniform correction to $\fsig$. Whether
this is correct at the accuracy level required remains to be
demonstrated with numerical simulations.

Our study illustrates just how daunting is the task of calibrating the
mass function to the accuracy of $\lesssim 5\%$.  Large numbers of
large-volume simulations are required to estimate the abundance of
cluster-sized objects, but high dynamic range is required to properly
resolve their internal mass distribution and subhalos. The numerical
and resolution effects should be carefully controlled, which requires
stringent convergence tests.  In addition, the abundance of halos on
the exponential cutoff of the mass function can be influenced by the
choice of method to generate initial conditions and the starting
redshift, as was recently demonstrated by \citet[][see also Appendix
A]{crocce_etal:06}. All this makes exhaustive studies of different
effects and cosmological parameters using brute force calibration of
the kind presented in this paper for the $\Lambda$CDM cosmology
extremely demanding. Clever new ways need to be developed both in the
choice of the parameter space to be investigated \citep{habib_etal07}
and in complementary studies of various effects using smaller,
targeted simulations.

\acknowledgements

\noindent 
We thank Roman Scoccimarro for simulations, computer time to analyze
them, and discussions on initial conditions. We thank Rebecca Stanek,
Gus Evrard, Martin White, and Uros Seljak for many helpful
discussions. We thank Alex Vikhlinin, Salman Habib, and David Weinberg
for useful discussions and comments on the manuscript. J.T. was
supported by the Chandra award GO5-6120B and National Science
Foundation (NSF) under grant AST-0239759.  A.V.K. is supported by the
NSF under grants No.  AST-0239759 and AST-0507666, by NASA through
grant NAG5-13274, and by the Kavli Institute for Cosmological Physics
at the University of Chicago.  Portions of this work were performed
under the auspices of the U.S. Dept. of Energy, and supported by its
contract \#W-7405-ENG-36 to Los Alamos National Laboratory.
Computational resources were provided by the LANL open supercomputing
initiative. S.G. acknowledges support by the German Academic Exchange
Service.  Some of the simulations were performed at the Leibniz
Rechenzentrum Munich, partly using German Grid infrastructure provided
by AstroGrid-D. The GADGET SPH simulations have been done in the
MareNostrum supercomputer at BSC-CNS (Spain) and analyzed at NIC
J\"ulich (Germany).  G.Y. and S.G. wish to thank A.I. Hispano-Alemanas
and DFG for financial support. G.Y. acknowledges support also from
M.E.C. grants FPA2006-01105 and AYA2006-15492-C03.


\break
\bibliography{risa}



\begin{figure*}
\epsscale{1.0} 
\vspace{-2.0cm}
\plotone{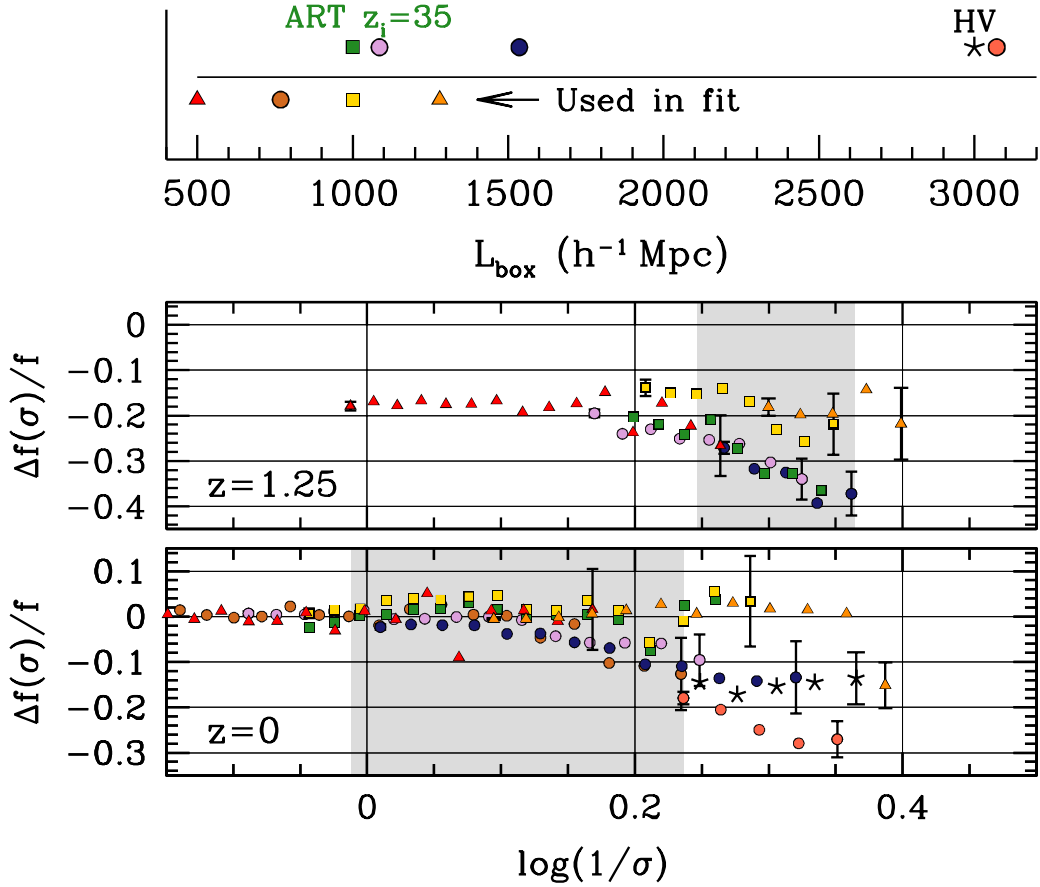}
\caption{ \label{appendix} Comparison between the large-box
  simulations used in the text and those in \cite{warren_etal:06} and
  \cite{evrard_etal:02}. The box sizes and point-types for the three
  HOT boxes and the Hubble Volume are shown in the top panel above the
  horizontal line. In addition, a version of the L1000W ART box,
  started at lower redshift, is also included in the comparison. The
  large-box simulations used from Table 1 are also included below the
  horizontal line. The bottom panel compares the $\D=200$ mass
  functions to the best-fit $\fsig$ at $z=0$. The middle panel shows
  the results at $z=1.25$. In the $z=0$ panel, the shaded region
  indicates $10^{14}$\hmsol\ $<M< 10^{15}$ \hmsol\ in the WMAP1
  cosmology. In the $z=1.25$ panel, the shaded region indicates
  $10^{14}$\hmsol\ $<M< 10^{14.5}$ \hmsol.}
\end{figure*}

\vspace{2cm}
\appendix
\section{A. Tests of the Initial Conditions}

In a recent study, \cite{crocce_etal:06} investigated differences
between using the standard first-order Zel'dovich Approximation (ZA)
and second-order Lagrangian perturbation theory (2LPT) for generating
initial conditions of cosmological simulations. ZA assumes that
particle trajectories are straight lines, but for large density
fluctuations trajectories should curve due to tidal effects. Thus, if a
simulation is initialized at the epoch where the overdensity is large
in some regions, the resulting error in particle trajectories will
lead to `transients' in the evolution of perturbations (see also
\citealt{roman:98}), which can persist to $z=0$. This effect is
strongest for the regions containing rarest peaks of largest height
that tend to evolve into the largest galaxy clusters at low redshift.
In their simulation results, Crocce et.~al.\ find a $\sim 10\%$
discrepancy at $M\sim 10^{15}$ \hmsol\ in $z=0$ mass functions
between 2LPT and ZA with starting redshift of $z_i=24$. This
discrepancy is expected to grow more significant at higher redshift at
fixed halo mass. The effect is particularly worrisome for
precision calibration of abundance of the most massive objects at any
redshift (those objects that are currently collapsing or have only
recently collapsed). In this appendix we present tests of the effects
of the initial redshift on the mass function and explain why we have
discarded some of the large-volume simulations from our analysis.

The top panel in Figure~\ref{appendix} shows a graphical key of the
three large-box HOT simulations used in the Warren et.~al.\ fit that
we do not utilize in our mass function fits. These simulations have
starting redshifts of $z_i=34$, 28, and 24 (with $z_i$ decreasing with
increasing box size). In addition, we also have results from the
Hubble Volume (HV) simulation, a $3000$ \hmpc\ simulation
(\citealt{evrard_etal:02}). We use the same SO halo catalog presented
in \cite{evrard_etal:02}, which used a density criterion of 200 times
the critical density rather than the mean. Thus we have scaled the
halo masses from $\D=666$ to $\D=200$, assuming NFW profiles as
detailed in \S 2. Lastly, we have included a re-simulation of the
L1000W ART box which has been initialized at $z_i=35$ rather than
$z_i=60$ using the same set of random phases and ZA at both starting
redshifts.

The bottom panels of Figure~\ref{appendix} show the residuals of the
simulation mass function from the best fit to our core simulation set
at $z=0$ and $z=1.25$.  At $10^{14}$ \hmsol, all simulations are in
excellent agreement. However, at $10^{15}$ \hmsol, the HOT boxes are
$\sim 10-20\%$ below the $\fsig$ obtained from the 2LPT simulations
and ART L1000W run.  The mass function of the HV simulation, with
$z_i=35$, is also $\sim 15\%$ below the 2LPT simulations.

At $z=0$, there is a $\sim 2\%$ difference between low-$z_i$ ART box
and the higher-$z_i$ version used in the fitting. This is smaller than
the difference between mass functions for the \citealt{crocce_etal:06}
simulations with $z_i=24$ and $z_i=49$, which may be due to sample
variance. However, the difference between the two ART boxes increases
at higher redshift. The ART box with $z_i=60$ is in good agreement
with the 2LPT simulations at $z=1.25$, implying that convergence has
been reached at a lower $z_i$ than shown in Crocce et~al. The run with
$z_i=35$, however, is $20-40\%$ lower than the best fit at large
masses.

It is not yet entirely clear whether the source of the discrepancies
in the mass functions at the highest masses can be attributed solely
to the errors of the ZA-generated initial conditions. The difference
between the large-volume HOT boxes and the 2LPT results are larger
than expected from just the ZA errors. Also, both ART boxes, with
$z_i=35$ and $z_i=60$, are in agreement with the 2LPT simulations at
$z=0$. Other factors, such as resolution effects on the halo density
profiles, may play a dominant role in the discrepancy exhibited by
both the HOT boxes and the HV simulation. Regardless of the source of
the discrepancy, it is clear that the large-volume HOT boxes and HV
simulations are systematically different from other higher-resolution
simulations. We therefore do not include them in our analyses.

In summary, the simulations which we use to derive our constraints on
the high-mass end of the halo mass function are all robust against
changing initial redshift. The 2LPT simulations have been thoroughly
tested in \citet{crocce_etal:06}. The L1000W and L500 simulations,
utilizing ZA with $z_i\gtrsim 50$, show consistent results with the
2LPT simulations at multiple redshifts. However, quantifying the
effects of initial conditions, finite simulation volume, and possible
numerical artifacts at the $\lesssim 1\%$ level will require
significant additional work.

\section{B. Interpolation of Mass Function Parameters}

To facilitate the use of our results in analytic calculations, 
we provide fitting functions for the
parameters of $\fsig$ as a function of $\log\D$. The dependence of
each parameter in the mass function is reasonably well described by

\begin{equation}
A = \left\{ \begin{array}{ll}
      0.1(\log\D) - 0.05 & {\rm if\ \ } \D< 1600 \\ 
      0.26 & {\rm if\ \ } \D\ge 1600,\\
      \end{array}
      \right. 
\end{equation}

\begin{equation}
a = 1.43 + (\log\D-2.3)^{1.5},
\end{equation}

\begin{equation}
b = 1.0 + (\log\D-1.6)^{-1.5},
\end{equation}

\noindent and

\begin{equation}
c = 1.2 + (\log\D-2.35)^{1.6}.
\end{equation}

\begin{figure*}
\epsscale{0.9} 
\plotone{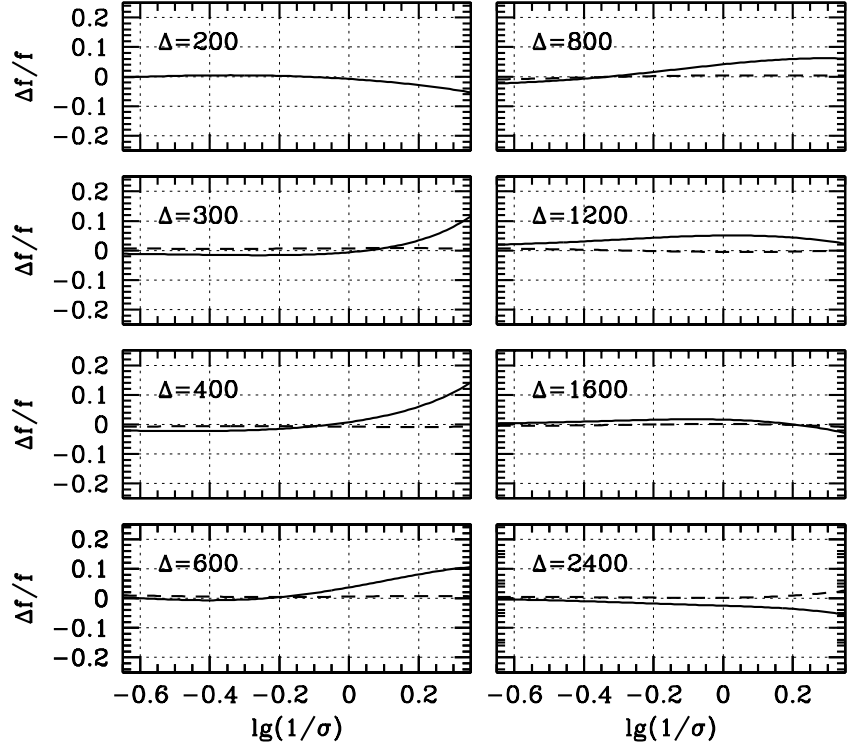}
\vspace{-2.0cm}
\caption{ \label{appB} Accuracy of the fitting functions presented in
  Appendix B for calculating the parameters of $\fsig$ as a function
  of $\D$ ({\it solid lines}). All curves are residuals with respect
  to the best-fit results of $\fsig$ from Table 2. For all
  overdensities except $\D=600$, the accuracy of $\fsig$ is $\lesssim
  5\%$. The dashed lines show the accuracy of $\fsig$ when using
  spline interpolation, which is accurate to $\lesssim 2\%$ for all
  $\D$ and $\siginv$. }
\end{figure*}

\noindent All logarithms are base 10. Because the parameters of
$\fsig$ are not completely smooth with $\log\D$, these functions yield
mass functions that are accurate to only $\lesssim 5\%$ for most
values of $\D$, but can degrade to $\lesssim 10\%$ at $\siginv>0.2$
for some overdensities. Figure \ref{appB} demonstrates the accuracy of
the fitting functions with respect to the results from Table 2. For
higher accuracy, we recommend spline interpolation of the parameters
as a function of $\log\D$. Figure \ref{appB} shows the results of the
spline interpolation when obtaining the parameters of $\fsig$. We
provide the second derivatives of the $\fsig$ parameters
for calculation of the spline coefficients (cf., \S 3.3 in
\citealt{press_etal:92}) in Table B3.

\begin{deluxetable*}{rrrrr}
\tablecolumns{5} 
\tablewidth{20pc} 
\tablecaption{Second derivatives of $f(\sigma)$ parameters}
\tablehead{\colhead{$\D$} & \colhead{$A$} & \colhead{$a$} & \colhead{$b$} & \colhead{$c$} }

\startdata

200 & 0.00 & 0.00 & 0.00 & 0.00 \\
300 & 0.50 & 1.19 & -1.08 & 0.94 \\
400 & -1.56 & -6.34 & 12.61 & -0.43 \\
600 & 3.05 & 21.36 & -20.96 & 4.61 \\
800 & -2.95 & -10.95 & 24.08 & 0.01 \\
1200 & 1.07 & 2.59 & -6.64 & 1.21 \\
1600 & -0.71 & -0.85 & 3.84 & 1.43 \\
2400 & 0.21 & -2.07 & -2.09 & 0.33 \\
3200 & 0.00 & 0.00 & 0.00 & 0.00 \\
\end{deluxetable*}

\section{C. An Alternate, Normalized Fitting Function}

The fitting function given in equation (\ref{e.fsig}) is an excellent
descriptor of the data over the range of our data, but at
$\siginv\lesssim 0.1$, $\fsig$ asymptotes to a constant value. For some
applications, specifically halo model calculations of dark matter
clustering statistics, it is necessary to integrate over all $\siginv$
to account for all the mass in the universe. Integrating
(\ref{e.fsig}) over all $\siginv$ yields infinite mass. In this
appendix we present an alternative fitting function that is normalized
such that

\begin{equation}
\label{e.norm}
\int \gsig \,d\,\ln \siginv = 1
\end{equation}

\noindent for all values of $\D$ at $z=0$. We focus on equation
(\ref{e.fsig}) for our main results because the parameters of that
function vary more smoothly and monotonically with $\D$, and
incorporating redshift evolution into that function is more
straightforward and more accurate. Because we can only calibrate our
mass function to $\siginv \gtrsim 0.25$, the behavior of the fitting
function at lower masses is arbitrary. Thus it is not to be expected
that the fitting function in this appendix is more or less accurate
than equation (\ref{e.fsig}) below this calibration limit, merely that
the function is better behaved.

With these caveats in mind, we find that at $z=0$ a function of the
form

\begin{equation}
\label{e.fsig_norm}
\gsig = B\left[\left(\frac{\sigma}{e}\right)^{-d} + \sigma^{-f}\right]e^{-g/\sigma^2}
\end{equation}

\noindent yields nearly identical results to those presented in Figure
\ref{diff_wmap1}. Equation (\ref{e.fsig_norm}) has four free
parameters, with $B$ set by the normalization constraint from
equation (\ref{e.norm}). We follow the same procedure for fitting the
model to the data as in \S 2.4. Best-fit parameters are listed in
Table C4. The $\xdof$ values are similar to the values listed in Table
2. The asymptotic slope of $\fsig$ in the \cite{sheth_tormen:99}
fitting function is $\sigma^{-0.4}$ at low masses. The values of $f$
in Table C4 vary around this value, with the $\D=200$ $\gsig$ going as
$\sigma^{-0.51}$ and $\D=3200$ going as $\sigma^{-0.33}$.

Another requirement of the halo model is that dark matter be unbiased
with respect to itself. This requires a recalibration of the
large-scale halo bias function, which we investigate in another paper
(Tinker et al., in preparation).

\begin{deluxetable*}{ccccccc}
\tablecolumns{7} 
\tablewidth{15pc} 
\tablecaption{Normalized Mass Function Parameters for $\gsig$ at $z=0$}
\tablehead{\colhead{$\D$} & \colhead{$B$} & \colhead{$d$} & \colhead{$e$} &\colhead {$f$} & \colhead{$g$} & \colhead{$\xdof$} }
\startdata

200 &  0.482 & 1.97 & 1.00 & 0.51 & 1.228 & 1.14 \\
300 &  0.466 & 2.06 & 0.99 & 0.48 & 1.310 & 1.16 \\ 
400 &  0.494 & 2.30 & 0.93 & 0.48 & 1.403 & 1.04 \\ 
600 &  0.494 & 2.56 & 0.93 & 0.45 & 1.553 & 1.07 \\ 
800 &  0.496 & 2.83 & 0.96 & 0.44 & 1.702 & 1.09 \\ 
1200 & 0.450 & 2.92 & 1.04 & 0.40 & 1.907 & 1.00 \\ 
1600 & 0.466 & 3.29 & 1.07 & 0.40 & 2.138 & 1.07 \\ 
2400 & 0.429 & 3.37 & 1.12 & 0.36 & 2.394 & 1.12 \\ 
3200 & 0.388 & 3.30 & 1.16 & 0.33 & 2.572 & 1.14 \\ 

\enddata
\end{deluxetable*}

\end{document}